\documentclass{elsarticle}
\usepackage[T1]{fontenc}
\usepackage[utf8]{inputenc}
\usepackage{lineno, hyperref}
\usepackage{longtable}
\usepackage{gensymb}
\usepackage{graphicx}
\usepackage{amsmath,amsfonts,amssymb}
\usepackage{rotating}
\usepackage{lscape}
\usepackage{natbib}
\usepackage{xcolor}
\usepackage[normalem]{ ulem }
\usepackage{soul}
\usepackage{cancel}
\modulolinenumbers[1]

\journal{Journal of \LaTeX\ Templates}

\newcommand{\cm}{cm$^{-1}$}
\newcommand{\dinitrogen}{N$_2$}
\newcommand{\methane}{CH$_4$}
\newcommand{\acetylene}{C$_2$H$_2$}
\newcommand{\ethane}{C$_2$H$_6$}
\newcommand{\propane}{C$_3$H$_8$}
\newcommand{\methyl}{C$_3$H$_4$}
\newcommand{\diacetylene}{C$_4$H$_2$}
\newcommand{\cyano}{HC$_3$N}
\newcommand{\benzene}{C$_6$H$_6$}
\newcommand{\dioxide}{CO$_2$}
\newcommand{\ethylene}{C$_2$H$_4$}

\newcommand\cites[1]{\citeauthor{#1}'s\ (\citeyear{#1})}
\begin{document}

\begin{frontmatter}

\title{Seasonal changes in the middle atmosphere of Titan from Cassini/CIRS observations: temperature and trace species abundance profiles from 2004 to 2017}

\author[1]{Christophe Mathé\thanks{christophe.mathe@obspm.fr}}
\author[1]{Sandrine Vinatier}
\author[1]{Bruno Bézard}
\author[2]{Sébastien Lebonnois}
\author[3]{Nicolas Gorius}
\author[3]{Donald E. Jennings}
\author[4]{Andrei Mamoutkine}
\author[5]{Ever Guandique}
\author[2]{Jan Vattant d'Ollone}

\address[1]{LESIA, Observatoire de Paris, Université PSL, CNRS, Sorbonne Université, Université de Paris, 5 place Jules Janssen, 92195 Meudon, France}
\address[2]{Laboratoire de Météorologie Dynamique, Sorbonne Université, ENS, Université PSL, Ecole Polytechnique, Université Paris-Saclay, CNRS, France}
\address[3]{NASA/Goddard Space Flight Center, Code 693, Greenbelt, MD 20771, USA}
\address[4]{Department of Astronomy, University of Maryland, College Park, MD 20742, USA}
\address[5]{ADNET Systems, Inc., Bethesda, MD 20817, USA}
\begin{abstract}
%le fait que cassini soit fini on a une demie annee de donnees et la derniere calibration, et donc faire une revue de CIRS
The Cassini/Composite InfraRed Spectrometer (CIRS) instrument has been observing the middle atmosphere of Titan over almost half a Saturnian year. We used the CIRS dataset processed through the up-to-date calibration pipeline to characterize seasonal changes of temperature and abundance profiles in the middle atmosphere of Titan, from mid-northern winter to early northern summer all around the satellite. We used limb spectra from 590 to 1500 \cm\/ at 0.5-\cm\/ spectral resolution, which allows us to probe different altitudes. We averaged the limb spectra recorded during each flyby on a fixed altitude grid to increase the signal-to-noise ratio. These thermal infrared data were analyzed by means of a radiative transfer code coupled with an inversion algorithm, in order to retrieve vertical temperature and abundance profiles. These profiles cover an altitude range of approximately 100 to 600 km, at 10- or 40-km vertical resolution (depending on the observation). Strong changes in temperature and composition occur in both polar regions where a vortex is in place during the winter. At this season, we observe a global enrichment in photochemical compounds in the mesosphere and stratosphere and a hot stratopause located around 0.01 mbar, both linked to downwelling in a pole-to-pole circulation cell. After the northern spring equinox, between December 2009 and April 2010, a stronger enhancement of photochemical compounds occurred at the north pole above the 0.01-mbar region, likely due to combined photochemical and dynamical effects. During the southern autumn in 2015, above the South pole, we also observed a strong enrichment in photochemical compounds that contributed to the cooling of the stratosphere above 0.2 mbar ($\sim$300 km). Close to the northern spring equinox, in December 2009, the thermal profile at 74$\degree$N exhibits an oscillation that we interpret in terms of an inertia-gravity wave.
\end{abstract}

\end{frontmatter}

%\linenumbers
%============================================================================%
% PART: INTRODUCTION
%============================================================================%
\section{Introduction}
% Titan, Cassini, GCM prediction
The Cassini spacecraft has been exploring the Saturnian system between July 2004 and September 2017. During almost half a Saturnian year, it performed 127 close flybys  of Titan. The stratosphere and mesosphere of Titan are mainly composed of \dinitrogen (98.4\%) and \methane (1.5\%) with traces of dihydrogen, hydrocarbons, nitriles and oxygen compounds \citep{Bezard2014}. The two main compounds, N$_2$ and CH$_4$, are dissociated by the ultraviolet solar flux and by electrons from Saturn's magnetosphere, which leads to the formation of photochemical compounds and aerosols through complex pathways. One-dimensional photochemical models, coupling neutral and ion chemistry together with vertical eddy transport and condensation, have been developed by different groups \citep[see, e.g.][for the most recent ones]{Krasnopolsky2014, Loison2015, Dobrijevic2016, Vuitton2019}. Saturn's obliquity induces strong seasonal changes in Titan's atmosphere. In winter/summer seasons, general circulation models (GCM) predict a pole-to-pole circulation cell in the middle atmosphere with the ascending branch located at the summer pole and the descending branch at the winter pole \citep{Newman2011, Lebonnois2012, Lora2015}. A polar vortex, present during winter, vanishes in early spring. Around the equinoxes, GCMs predict the presence of two equator-to-pole circulation cells in the middle atmosphere, with an ascending branch at low latitudes and descending branches at both poles \citep{Newman2011, Lebonnois2012, Lora2015}. The descending branches of these pole-to-pole or equator-to-pole circulation cells enrich in photochemical compounds and adiabatically heats the mesosphere at high latitudes. 

% Background: CIRS
% Vinatier 2015 from 2006 to 2013
% Teanby 2017 formation polar vortex pole Sud
Various analyses of spectra recorded by the Cassini/CIRS (Composite InfraRed Spectrometer) have highlighted seasonal changes in temperature and abundance profiles (for the most recent publications, see e.g., [\cite{Vinatier2015, Teanby2017, Sylvestre2018, Coustenis2019, Teanby2019}]). In particular, thermal and abundance profiles in the middle atmosphere (between 20 and 0.001 mbar) have been monitored by \cite{Vinatier2015} from 2006 to 2013. In the northern polar region, the temperatures in the pressure range 0.4-0.002 mbar observed during the winter, from 2006 to 2008, were higher than in the 2010-2011 period. This probably results from the adiabatic heating provided by the descending branch of the pole-to-pole circulation cell which is stronger during winter \citep{Achterberg2011}. Still during winter, the northern polar region was enriched in photochemical compounds compared to the equatorial region, from the lower stratosphere \citep{Sylvestre2018, Coustenis2019, Teanby2019} to the mesosphere \citep{Vinatier2015}. Then, from 2009 to 2010, shortly after northern spring equinox (August 2009), the northern polar region exhibited a stronger enrichment in photochemical compounds above 0.1 mbar compared to winter. This feature may result from a combination of dynamics and chemistry, in which the solar flux reappearing after the polar night locally reactivates the photochemical production of most species that are then transported downwards by the equator-to-pole circulation cell \citep{Vinatier2015}. This study also suggests that both equator-to-pole cells coexisted from at last January 2010 to at least June 2010. In the southern polar region, a mesospheric hot spot around 0.01 mbar ($\sim$400 km) appeared between 2010 and 2011, likely due to the increasing strength of the adiabatic heating provided by the descending air branch \citep{Teanby2012, Teanby2017, Vinatier2015}. During northern spring, between 2012 and early 2015, \cite{Teanby2017} observed a cold mesosphere near the south pole, which they attributed to a radiative effect linked to the mesospheric enrichment in photochemical compounds inside the polar vortex. This trace gas enrichment, likely resulting from downwelling in the pole-to-pole circulation cell, was also observed in the lower stratosphere at 15 mbar \citep{Sylvestre2018}. From 2015 onwards, the mesosphere near the south pole warmed up, probably due to the increase of the subsidence velocity in the descending branch of the circulation cell \citep{Teanby2017}.
 
% Sylvestre 2018 lower stratosphere FP1
%In the lower stratosphere, \cite{Sylvestre2018} focus on the seasonal evolution of cyanogen (C$_2$N$_2$), methylacetylene (\methyl) and diacetylene (\diacetylene) during the entire Cassini mission. They show no seasonal changes of these species at high northern latitudes from 2004 to 2016. In contrast, at high southern latitudes, these species were suddenly enriched after 2012, during the southern autumn.

% Coustenis 2018 nadir winter pole Temperature et FP3 FP4
%\cite{Coustenis2018} analyzed nadir spectra with 0.5 \cm spectral resolution, at winter poles, in order to constrain the lower stratosphere ($\sim$ 100 km). The authors detected benzene at the northern winter pole.

\cite{Vinatier2015} studied the seasonal changes between October 2006 and May 2012 using CIRS mid-IR limb spectra acquired at the highest available spectral resolution (0.5 \cm). These authors inferred thermal and abundance profiles (C$_2$H$_2$, C$_2$H$_4$, C$_2$H$_6$, C$_3$H$_8$, C$_3$H$_4$, C$_4$H$_2$, C$_6$H$_6$, CO$_2$, HCN, HC$_3$N) at altitudes between 100 and 550 km. In this study, we extended the analysis of the 0.5 \cm \/ mid-IR limb spectra to the whole dataset acquired during the entire Cassini mission between 2004 and 2017, and using a new methodology, we were able to probe from 100 to around 650 km altitude.  This study is complementary to the work of \cite{Coustenis2019} who analyzed nadir CIRS mid-IR spectra at 0.5 \cm \/ spectral resolution between 2010 and 2017, probing deeper altitude levels (between 20 mbar and 0.05 mbar, depending on the molecule, i.e. $\sim$ 80 and 315 km). \cite{Teanby2019} analyzed mid-IR CIRS nadir spectra acquired with a lower spectral resolution of 2.5 \cm \/ between 2004 and 2017, probing similar altitude levels (5 to 0.1 mbar, $\sim$ 115 to 280 km) than \cite{Coustenis2019}. Using CIRS far-IR nadir spectra, \cite{Sylvestre2018} monitored seasonal changes in C$_4$H$_2$, C$_3$H$_4$ and C$_2$N$_2$ abundance profiles at 15 mbar ($\sim$ 85 km), which is deeper than levels probed in the mid-IR spectral range.

% Motivation
We present here an analysis of the whole set of Cassini/CIRS limb spectra recorded at 0.5-\cm\ resolution during 56 of the 127 Titan flybys. Our goal is to document the seasonal changes in temperature and composition that take place in the stratosphere and lower mesosphere, at all covered latitudes (poles, mid-latitudes and equator), and during the entire Cassini mission from 2004 to 2017. Data selection is described in Section 2 and the methodology used to retrieve the temperature and gas abundance profiles is detailed in Section 3. Thermal and abundance profiles of C$_2$H$_2$, C$_2$H$_4$, C$_2$H$_6$, C$_3$H$_4$, C$_3$H$_8$, C$_4$H$_2$, C$_6$H$_6$, HCN, HC$_3$N and CO$_2$ are shown in Section 4. The results are discussed in Section 5 and a brief conclusion is presented in Section 6.

%============================================================================%
% PART: OBSERVATIONS
%============================================================================%
\section{Observations}
% Description CIRS & geometry observation
The CIRS spectrometer was composed of three focal planes (FP) covering the 10-1500 \cm \space wavenumber range: FP1 (10 to 600 \cm), FP3 (580 to 1100 \cm) and FP4 (1050 to 1500 \cm). We used both FP3 and FP4, which are each composed of an array of 1$\times$10 pixels. This linear shape is relevant for limb viewing geometry, as the line-of-sight of each pixel traverses the atmosphere down to a given tangent height, which varies from the surface to about 700 km. However, even for low tangent heights, limb spectra cannot probe deeper than a pressure level of $\sim$5-6 mbar ($\sim$115 km), due to the large increase of the opacity in the molecular emission bands around this pressure level. Each pixel has a 0.27$\times$0.27 mrad$^2$ field-of-view corresponding to a vertical resolution varying from 10 to 40 km (comparable to a pressure scale height), depending on the distance of the spacecraft to Titan during the flyby. More details on the CIRS instrument are given in \cite{Kunde1996}, \cite{Flasar2004} and \cite{Jennings2017}, and more details on the different types of observations are given in \cite{Nixon2019}.

% Details on spectra selection
We used the CIRS ``global calibration'' database described in \cite{Jennings2017}. We selected limb spectra with a spectral resolution of 0.5 \cm\ for which gas molecular emission bands are well separated. Spectral selections cover the entire Cassini mission, from the northern winter in December 2004 to the early northern summer in September 2017, and pole-to-pole latitudes. Characteristics of the observations that we used are listed in Table \ref{table:list_obs} (Appendix A).

% Linear combination
In order to increase the signal-to-noise ratio, we averaged all available limb spectra acquired during a given observation sequence and having close enough tangent heights (see Appendix B for more details on our averaging method). For some flybys (Tb, T06, T16, T28, T59, T84, T103, T120), we complemented the limb spectra with an average of nadir spectra to extend the information available from the retrievals to lower altitudes.

% Correction on spectra
For a couple of observations, limb spectra with the highest tangent heights presented negative continua in the 1075-1200 \cm\ spectral range due to calibration problems. We developed a procedure that corrects this negative continuum, and allowed us to determine the error due to this effect on temperature and abundance profile retrievals (see Appendix C).

%============================================================================%
% PART: METHODOLOGY
%============================================================================%
\section{Retrieval methodology}
CIRS acquired spectra of the thermal emission of Titan's atmosphere. Molecular gases emit through their rotation and ro-vibration bands, the intensity of which depends on both temperature and gas mixing ratio profiles. The continuum emission is due to the haze opacity and collision-induced absorption of N$_2$-N$_2$, N$_2$-H$_2$, N$_2$-CH$_4$ and CH$_4$-CH$_4$.

To infer thermal and molecular mixing ratio profiles from CIRS observations, we proceeded as follows: first, we retrieved the thermal profile by fitting the $\nu_4$ CH$_4$ band at 1306 \cm, assuming a constant methane mixing ratio of 1.48\%, as measured \textit{in situ} by the GCMS instrument \citep{Niemann2010}. The impact of this hypothesis on the retrieved thermal profiles is discussed in Appendix D. We also assumed that temperature profiles did not vary with longitude as suggested by \cite{Flasar2005}. Secondly, molecular gas mixing ratio profiles were inferred using the retrieved thermal profile. Each profile retrieval have been performed using a correlation length equal to the vertical resolution of the limb spectra dataset for a given observation (see table \ref{table:list_obs}).

%\subsection{Code description}
Retrievals were performed using a line-by-line radiative transfer code coupled with a constrained linear inversion algorithm, described in \cite{Conrath1998} and \cite{Vinatier2015}.

%\subsection{Update on the code}
%With respect to \cite{Vinatier2015}, an update has been done in the inversion algorithm part: matrix inversion is performed by Cholesky decomposition in order to reduce computational time. The subroutine is adapted from Numerical recipes in Fortran Second Edition \citep{Press1996}. 
References for the spectroscopic data used here can be found in \cite{Vinatier2010a, Vinatier2015}, except for the HCN, HC$^{15}$N and H$^{13}$CN spectroscopic files that we updated from GEISA \citep{Jacquinet-Husson2011}. We also enlarged the spectroscopic dataset of C$_3$H$_8$ by incorporating the pseudo line-lists of the $\nu_7$ band (1100-1200 \cm) and the $\nu_{21}$ band (900-940 \cm) from \cite{Sung2013}.

\subsection{Thermal profile retrievals}
\label{methodology: thermal}
\begin{sidewaysfigure}
\includegraphics[width=\textwidth]{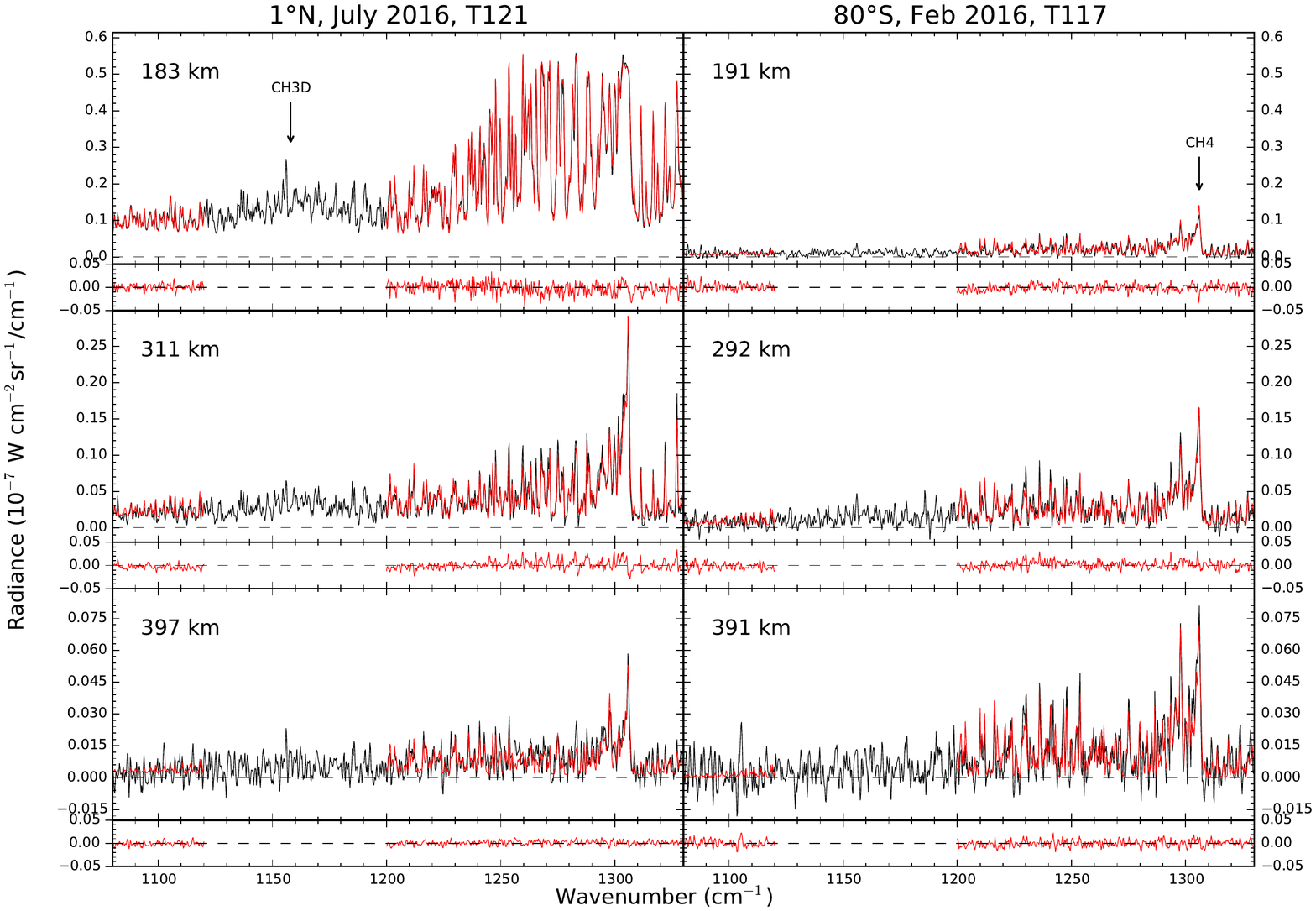}
\caption{FP4 limb spectra near the equator (left column) and south pole (right column), during northern spring at similar altitudes. Black lines correspond to the observed spectra and red lines correspond to the synthetic ones calculated in the region used in the inversion process. Residual radiances (observed - synthetic) are also plotted below. The altitudes correspond to the distance of the line-of-sight to the surface corrected by the vertical altitude shift. Molecules are indicated at the center of their bands.}
\label{fig: spectra_fp4}
\end{sidewaysfigure}

Temperature and haze optical depth vertical profiles were simultaneously retrieved. We derived the haze optical depth from the fit of the observed continuum in the 1080-1120 \cm\/ spectral range, since this spectral range is free of molecular band emission. The temperature profile was derived from the fit of the $\nu_4$ CH$_4$ band in the 1200-1330 \cm\/ spectral range (Fig.\ \ref{fig: spectra_fp4}). The \textit{a priori} temperature profile was the one derived at the equator (from the Tb flyby observations) by \cite{Vinatier2007a}.

% Spatial shift
We first determined the vertical shift to apply on the nominal altitude of the line-of-sight extracted from the CIRS database. This nominal altitude usually does not allow us to reproduce perfectly the relative intensity of the Q- and R- or P-branches of the $\nu_4$ CH$_4$ band. The difference between the nominal and the physical altitude is due to: (i) CIRS navigation pointing errors and (ii) our calculated pressure/altitude grid derived from hydrostatic equilibrium incorporating the \textit{in situ} temperature profile measured by HASI \citep{Fulchignoni2005} below 120 km, while CIRS observations probe at higher altitudes. So, we ran retrievals of our deepest averaged spectra with different vertical shifts applied to the altitudes of their line-of-sights to infer the one that minimizes the root mean square (rms) residuals between calculated and observed limb spectra (see Appendix B).

% Spectral shift
After having set the altitude shift, we determined the spectral shift to apply to the observed spectra to obtain the best fit of the 1198-1330 cm$^{-1}$ spectral range, covering the P-, Q- and part of R-branch of the $\nu_4$ CH$_4$ band. During the Cassini mission, the CIRS reference wavelength laser changed mode about 15 times and the CIRS wavelength nominal calibration was performed by measuring the position of emission lines of methane near 1330 \cm\/ for each mode. The spectral shift was then estimated to be less than 0.01 \cm\/ at 1300 \cm\/ \citep{Jennings2017, Brasunas2012}. We ran retrievals with different spectral shifts and kept the one that minimized the rms residuals between calculated and observed limb spectra. We noticed from our extracted datasets that a slight spectral shift of 0.02 \cm\/ remained, which could be due to the thermal drift of the reference wavelength laser diode. We therefore systematically determined the spectral shift from FP3 and FP4 limb observations independently, since the spectral shift varies with wavenumber.
\newline
%\textcolor{red}{\sout{After having set the altitude shift, we determined the spectral shift to apply to the observed spectra to obtain the best fit of the 1198-1330 \cm\/ spectral range, covering the P-, Q- and part of R-branch of the $\nu_4$ CH$_4$ band. During the Cassini mission, the CIRS reference wavelength laser changed mode about 15 times and the CIRS wavelength nominal calibration was performed by measuring the position of emission lines of methane near 1330 \cm\/ for each mode. The spectral shift was then estimated to be less than 0.01 \cm\/ at 1300 \cm\/ \citep{Jennings2017, Brasunas2012}. However, we noticed from our extracted datasets  that a slight spectral shift of 0.02 \cm\/ remained, which could be due to the thermal drift of the reference wavelength laser diode. We therefore systematically determined the spectral shift from FP3 and FP4 limb observations independently, since the spectral shift varies with wavenumber.}}

% Error bar calculation
The 1-$\sigma$ error on temperature were derived from the contribution of both the propagation of the spectral noise in the retrievals ($\Delta T_{noise}$) and the error ($\Delta T_{\Delta km}$) due to the uncertainty on the applied vertical shift ($\Delta z$, see below). The spectral noise contribution to the error bars ($\Delta T_{noise}$) was calculated from the covariance matrix of the solution using a tabulated noise equivalent spectral radiance (NESR), based on early observations during the mission, divided by the square root of the number of spectra as the measurement error. However, the actual NESR varies with the observing sequence and is usually lower than the rms residuals between calculated and observed limb spectra. So we scaled the temperature error $\Delta T_{noise}$ due to propagation of the noise by a factor of max(rms)/max(NESR), where max(rms) is the maximum value among all limb spectra of the rms residuals over the 1198-1330 \cm \/ interval and max(NESR) is the NESR value at 1330 \cm. Uncertainty on the vertical shift was determined from the rms of the residuals between observed and calculated spectra using different altitude shift values. The 1-sigma variation of the rms of the residual around the smallest value (corresponding to our bet fit) typically corresponds to a $\Delta z = \pm 2$ km variation. This uncertainty is the same as that determined by \cite{Vinatier2010b} and we verified that it was also valid for our dataset.

%The 1-$\sigma$ error on temperature and haze extinction profiles were derived from the propagation of the spectral noise in the retrievals and the error uncertainty on the vertical applied shift ($\delta T_{shift}$), which is typically of $\pm$2 km. The spectral noise contribution to the error bars was calculated from the covariance matrix of the solution using a tabulated noise equivalent spectral radiance \textcolor{red}{(NESR)}, based on early observations during the mission, divided by the square root of the number of spectra \textcolor{red}{\sout{(NESR)}} as the measurement error. However, the actual NESR varies with the observing sequence and is usually lower than the rms residuals between calculated and observed limb spectra. So we scaled the temperature error $\delta T_{noise}$ due to propagation of the noise by a factor of max(rms)/max(NESR), where max(rms) is the maximum value of the rms residual for all limb spectra used in the inversion process and max(NESR) is the value at 1330 \cm. The total 1-$\sigma$ error on the temperature profile is then:

\begin{equation}
\Delta T = \sqrt{(\Delta T_{noise})^2 \times \left(\frac{max(rms)}{max(NESR)}\right)^2 + (\Delta T_{\Delta km})^2}
\end{equation}

For observations in December 2006 (17$\degree$N, T21) and January 2007 (4$\degree$N, T23), highest limb spectra presented negative continua that we corrected through a procedure described in Appendix C. We added the error contribution of this correction that we estimated as the temperature difference between the retrievals for spectra with negative continuum and those for the corrected spectra. We found a maximum temperature difference of 5 K at high altitude (usually for tangent heights higher than 400 km) where corrupted limb spectra are most easily observed.

% Test of the robustness
We also assessed the information content of our retrieved temperature profiles by using the output profile, modified by typically 5-10 K, as a new input to the retrieval process. We then defined the region of validity of our retrievals as the altitude range where they do not significantly vary with the assumed \textit{a priori} profile.

\subsection{Molecular gas volume mixing ratio profile retrievals}
\label{methodology: molecule}
\begin{sidewaysfigure}
\includegraphics[width=\textwidth]{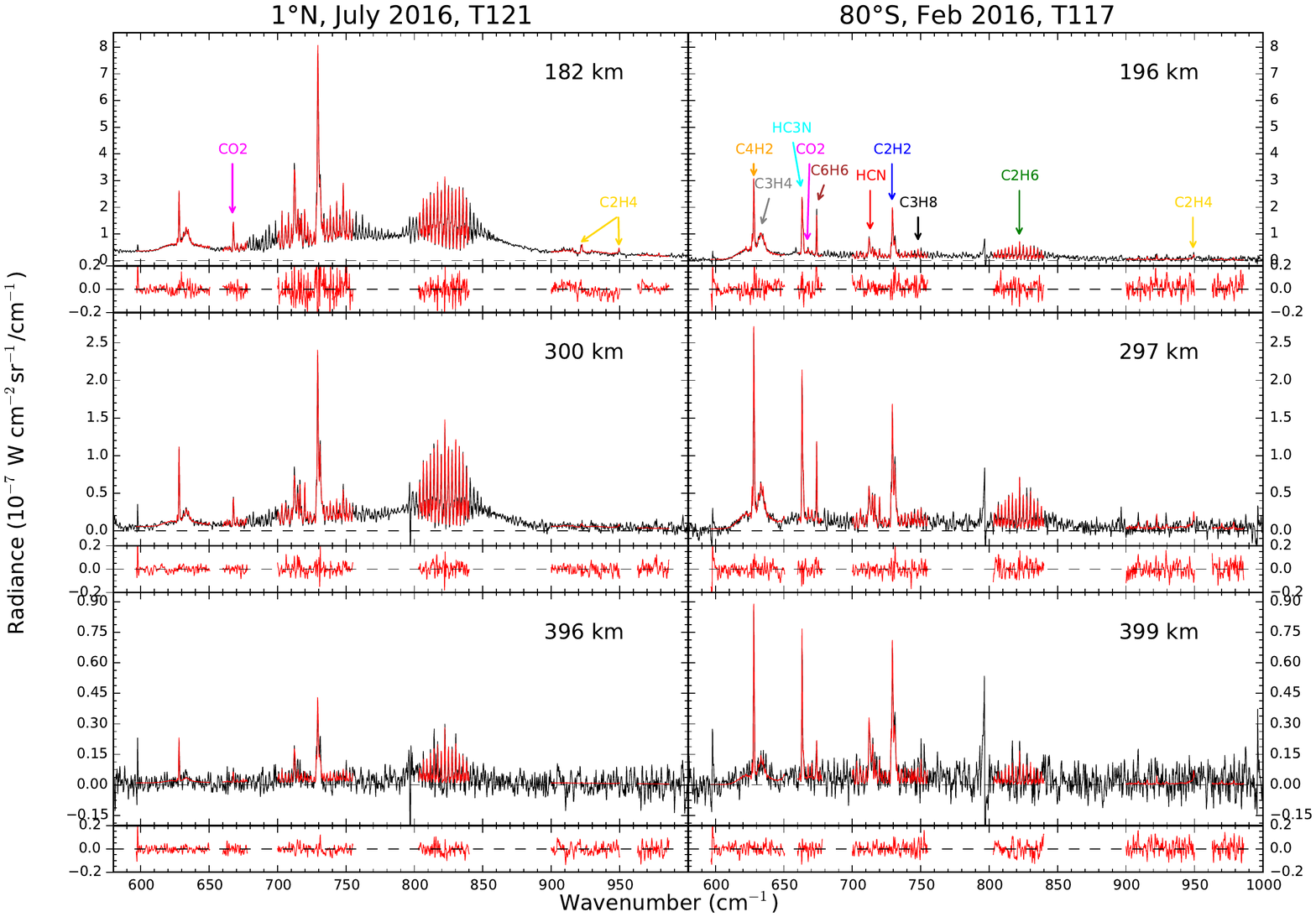}
\caption{Same as Fig.\ \ref{fig: spectra_fp4} for FP3 limb spectra near the equator (left column) and south pole (right column) during northern spring at similar altitudes. The features close to 598 and 795 cm$^{-1}$ are due to instrumental noise.}
\label{fig: spectra_fp3}
\end{sidewaysfigure}

The molecules observed in the CIRS FP3 spectral range are: C$_2$H$_2$, C$_2$H$_4$, C$_2$H$_6$, C$_3$H$_8$, C$_3$H$_4$, C$_4$H$_2$, C$_6$H$_6$, HCN, HC$_3$N and CO$_2$ (Fig.\ \ref{fig: spectra_fp3}). The spectral shift was determined from the fit of the 700-720 \cm\/ spectral range, which covers the P- and Q-branches of the $\nu_5$ C$_2$H$_2$ band and the Q-branch of the $\nu_2$ HCN band. The vertical shifts inferred from the FP4 limb spectra were applied to the FP3 limb spectra. 
We tested the robustness of the retrieved gas mixing ratio profiles by using different \textit{a priori} profiles. We first used constant-with-height profiles with volume mixing ratio spanning two to three orders of magnitude. Then, we ran a last test using an \textit{a priori} mixing ratio profile corresponding to the convergence zone of the previous retrieved profiles, and above the convergence zone we fixed the profile to a constant value. We finally retained as the region of validity the altitude range where the retrievals do not significantly depend on the \textit{a priori} profile. The haze extinction vertical profile retrieved from the FP4 spectra was used as the \textit{a priori} profile in the retrieval of the haze extinction profile from the FP3 spectral range, taking into account the spectral dependency of the haze opacity derived from \cite{Vinatier2012}.

Molecular gas mixing ratio and haze extinction profiles were retrieved simultaneously in a given spectral range. We retrieved in a first step the mixing ratio profiles of C$_2$H$_2$ ($\nu_5$ band at 729 \cm), HCN ($\nu_2$ band at 712 \cm) and the haze extinction in the 700-740 \cm \/ spectral range, since C$_2$H$_2$ has the most intense emission band in the FP3 spectral range. Then, we retrieved C$_2$H$_6$ ($\nu_9$ band at 822 \cm), C$_3$H$_8$ ($\nu_{26}$ band at 748 \cm) mixing ratios and haze extinction from the inversion of limb spectra in the 740-755 and 803-840 \cm \/ spectral ranges. Next, we retrieved C$_3$H$_4$ ($\nu_9$ band at 633 \cm), C$_4$H$_2$ ($\nu_8$ band at 628 \cm) mixing ratios and haze extinction from the 610-645 \cm \/ spectral range. HC$_3$N ($\nu_5$ band at 663 \cm), CO$_2$ ($\nu_2$ band at 667 \cm), C$_6$H$_6$ ($\nu_4$ band at 673 \cm) mixing ratios and haze extinction were retrieved simultaneously from the 660-678 \cm \/ spectral range. In a last step, we retrieved the C$_2$H$_4$ ($\nu_7$ band at 949 \cm) mixing ratio and haze extinction profiles from the 900-950 and 963-986 \cm\/ spectral ranges.

% Upper limit
For spectra in which C$_6$H$_6$ or HC$_3$N were not detected, we estimated the 2-$\sigma$ upper limit on their mixing ratios following the methodology described in \cite{Nixon2010}.

% Error bar calculation
The total relative error on the retrieved mixing ratio is the quadratic sum of the errors due to the spectral noise propagation $(\frac{\Delta q}{q})_{noise}$, the uncertainty on the altitude shift $(\frac{\Delta q}{q})_{\Delta km}$ and on the temperature profile $(\frac{\Delta q}{q})_{\Delta T}$ \citep{Vinatier2010b}:
\begin{equation}
\frac{\Delta q}{q} = \sqrt{ \left(\frac{\Delta q}{q}\right)^2_{noise} \times \left(\frac{max(rms)}{max(NESR)} \right)^2 + \left(\frac{\Delta q}{q}\right)^2_{\Delta km} + \left(\frac{\Delta q}{q}\right)^2_{\Delta T}}
\end{equation}
The uncertainty linked to the altitude shift $(\frac{\Delta q}{q})_{\Delta km}$ is the relative difference between the best-fit abundance profile and the abundance profile retrieved with the altitude shift ($\Delta z = \pm 2$ km) applied on the line-of-sight of limb spectra and using the temperature profile inferred with the same altitude shift. The relative uncertainty $(\frac{\Delta q}{q})_{\Delta T}$ is the relative difference between the best fit abundance profile and the one retrieved with the minimum or maximum thermal profile determined with $\Delta T$ derieved from Eq. 1. We determined these relative uncertainties for three cases: T35 (north), T47 (equator), T115 (south) and we considered that they do not vary over the entire studied time period.

%============================================================================%
% PART: RESULTS
%============================================================================%
\section{Results}
\subsection{Thermal profiles}
\begin{figure}
\includegraphics[width=\linewidth]{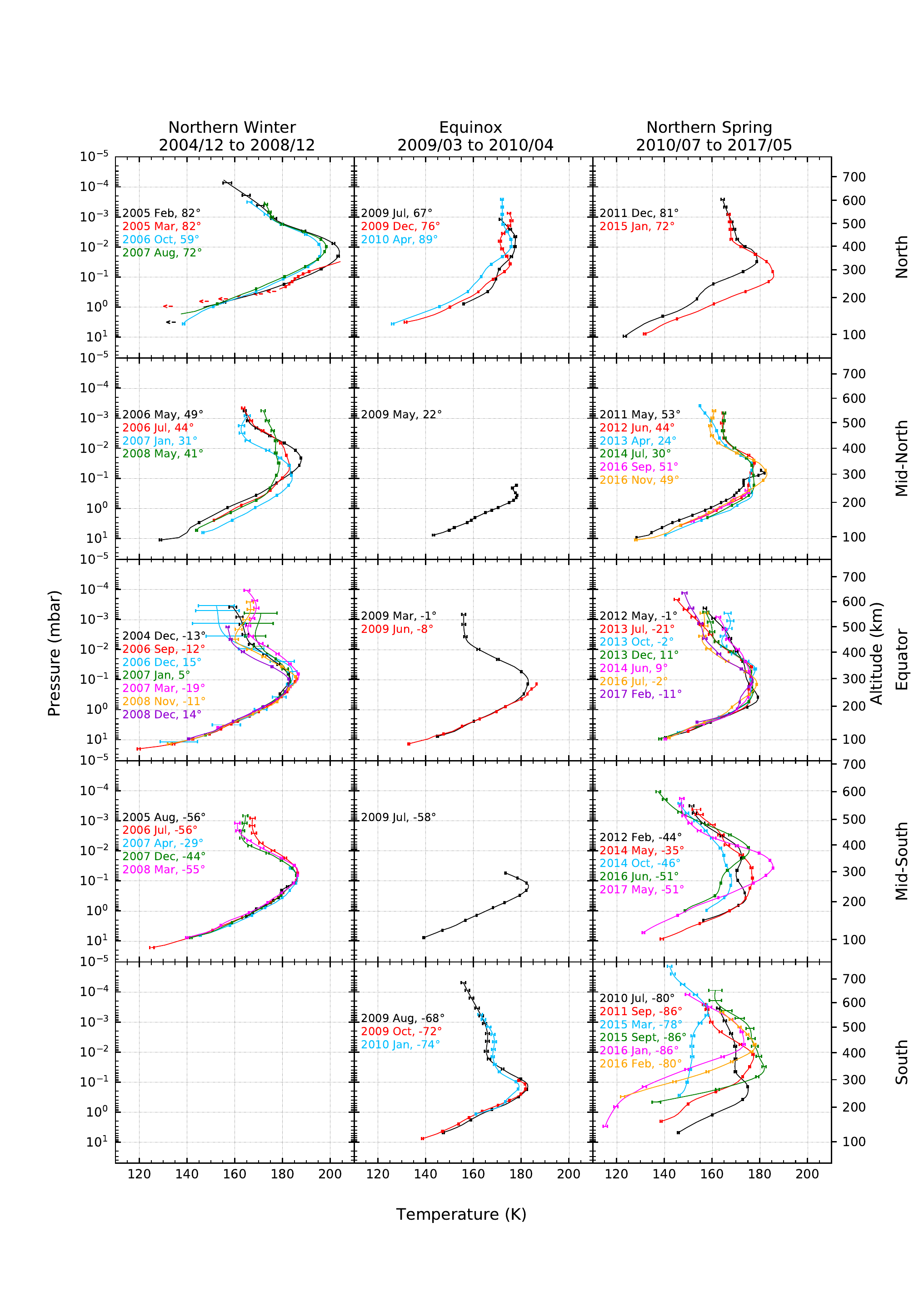}
\caption{Thermal profiles in the middle atmosphere of Titan. Each row corresponds to a given latitude region: from top to bottom, north pole to south pole. Each column corresponds to a season: left to right, northern winter to northern spring. On each thermal profile, the 1-$\sigma$ errors at the pressure level corresponding to the lines-of-sight of the spectra used for the retrievals are plotted. Arrows correspond to 2-$\sigma$ upper limits. The altitude grid is calculated using the temperature profile of: October 2006 for north, June 2012 for mid-north, May 2012 for equator, June 2016 for mid-south and October 2009 for south.}
\label{fig:fig_proft_1}
\end{figure}

Figure \ref{fig:fig_proft_1} shows the thermal profiles retrieved from the limb observations listed in Table \ref{table:list_obs}. These thermal profiles are gathered by latitude range, from top to bottom: north pole (90$\degree$N to 60$\degree$N), mid-north (60$\degree$N to 20$\degree$N), equator (20$\degree$N to 20$\degree$S), mid-south (20$\degree$S to 60$\degree$S) and south pole (60$\degree$S to 90$\degree$S) and by season, from left to right: northern winter (December 2004 to December 2008), northern spring equinox (March 2009 to April 2010) and northern spring (July 2010 to May 2017).

%We also refer to a specific temperature profile with molecular gas mixing ratio profiles corresponding to the same cell (see figures \ref{fig:fig_profq_c2h2} to \ref{fig:fig_profq_c2h4}).

\subsubsection{High northern latitudes}
During northern winter, the northern polar region above 0.1 mbar ($\sim$270 km) is hotter than the equatorial region, with a temperature maximum of $\sim$205 K at $\sim$10$^{-2}$ mbar ($\sim$400 km) in February 2005 at 82$\degree$N, which is the warmest temperature ever observed. In contrast, the northern polar region is colder than the equatorial region below 0.1 mbar ($\sim$270 km), e.g.\ at most 150 K at 1 mbar ($\sim$175 km) in 2005 at 82$\degree$N {\em vs} $\sim$170 K in the equatorial region. Near 4 mbar, the temperature in all four profiles is less than 140 K, more than 15 K lower than at equatorial latitudes.

From August 2007 to June 2009, i.e.\ near northern spring equinox, the northern polar region has been cooling in the 0.3--3$\times$10$^{-3}$ mbar ($\sim$310--430 km) region, by e.g.\ $\sim$23 K at 10$^{-2}$ mbar ($\sim$360 km). In December 2009 at 76$\degree$N, the thermal profile presented an oscillation between 4$\times$10$^{-2}$ mbar ($\sim$310 km) and 10$^{-3}$ mbar ($\sim$490 km) that is discussed in Section \ref{discussion: oscillation}. 

During northern spring, from December 2011 to January 2015, the northern polar region warmed up by $\sim$7 K between 10 mbar ($\sim$90 km) and 1 mbar ($\sim$170 km), and by $\sim$15 K between 1 mbar and 0.1 mbar ($\sim$280 km). On the other hand, above 0.02 mbar ($\sim$370 km), temperature did not significantly vary in this time lapse. In January 2015, the temperature maximum in the northern polar region was located around 0.1 mbar, two pressure scale heights lower than in 2005, and then similar to the location of the stratopause in the equatorial region.

%the stratopause became well defined in December 2011, with a temperature maximum of 180 K at 4*10$^{-2}$ mbar (400 km). In January 2015, the north pole was out of the polar hood and the solar flux in this region increased. The stratosphere warmed up, the stratopause is located at around , deeper than during the northern winter, with a stratopause temperature reaching 188 K.

\subsubsection{Northern mid-latitudes}
During northern winter, between May 2006 and May 2008, thermal profiles at latitudes poleward of 40$\degree$N were $\sim$5 K colder than that at 31$\degree$N between 7 mbar ($\sim$105 km) and 0.1 mbar ($\sim$280 km). The 31$\degree$N profile was similar to those in the equatorial region. In contrast, thermal profiles above 40$\degree$N were $\sim$10 K warmer above 0.02 mbar ($\sim$370 km) than the one at 31$\degree$N. At these latitudes, the temperature around 0.02 mbar decreased from $\sim$190 K in May 2006 to $\sim$177 K in May 2008.

Around northern spring equinox in May 2009, the temperature profile at 22$\degree$N was close to the profile at 31$\degree$N in January 2007 between 9 mbar ($\sim$100 km) and 0.5 mbar ($\sim$210 km), but colder in the 0.2-0.5 mbar region. These data are at high vertical resolution ($\sim$10 km) and do not extend higher than 0.1 mbar ($\sim$285 km). 

During northern spring, the stratosphere warmed up by about 3 K between 5 mbar ($\sim$120 km) and 0.3 mbar ($\sim$235 km) from May 2011 (53$\degree$N) to June 2012 (44$\degree$N). The temperature profile was roughly isothermal between 0.3 mbar and 0.04 mbar ($\sim$330 km) from June 2012 to July 2014 ($T \sim$175 K). In November 2016, the temperature had increased in this altitude region, reaching a maximum of 183 K at 0.08 mbar ($\sim$295 km).

\subsubsection{Equatorial latitudes}
Over the Cassini mission, the lower stratosphere, below the 0.5-mbar level, did not show temperature variations in excess of 5 K at equatorial latitudes. In the 0.1-mbar region, we do not observe strong variations between December 2004 and June 2009 ($T \sim$175 K), while in the 2012-2017 period, the temperature is $\sim$7 K lower. Higher, around 0.01 mbar, the situation is less clear but the temperature overall decreases by some 10 K between 2012 and 2017. In the $\mu$bar region, thermal profiles show variations over the mission that are correlated neither with latitude nor local time.

% On the other hand, the pressure level of the stratopause has changed during this period: (1) it was located at 10$^{-1}$ mbar (value in km) between December 2004 and June 2009, reaching 185 K in March 2007; (2) deeper in May 2016 at 5$\times$10$^{-1}$ mbar (value in km), (3) the disappearance of a well defined stratopause between July 2013 and June 2014, (4) the reappearance of the stratopause in July 2016 at around 1 mbar (value in km) reaching 178 K.

%During the northern winter, the stratosphere did not present strong variations in the equatorial region. The stratopause temperature decreased from $\sim$ 188 K during the northern winter to $\sim$ 179 K during the northern spring. This can be explain by the reduced solar flux due to the eccentricity of Saturn's orbit.
%However, the mesospheric temperature decreased by $\sim$ 15 K from December 2004 to December 2006 at 10$^{-3}$ mbar and then increased by 15 K by before January 2007, decreasing by 9 K b December 2008. In contrast, the mesospheric temp could be explained by the difference in latitude: coldest mesosphere are located in the northern hemisphere, while for warmest mesosphere are located in the southern hemisphere.

\subsubsection{Southern mid-latitudes}
During southern summer (2004-2008), temperature profiles at southern mid-latitudes did not present strong variations and were quite similar to those around the equator at the same period.
% The stratopause was located at $\sim$8$\times$10$^{-2}$ mbar (value in km), with a temperature around 185 K.

Near southern autumn equinox, in July 2009 at 58$\degree$S, the stratosphere had slightly cooled, by $\sim$2 K between 8 mbar ($\sim$110 km) and 0.07 mbar ($\sim$310 km), compared with southern summer conditions.

During southern autumn, the atmosphere in February 2012 (44$\degree$S) was $\sim$2 K warmer between 10 mbar ($\sim$100 km) and 1 mbar ($\sim$185 km) than that in July 2009 (58$\degree$S), while above the 1-mbar level the temperature was lower. The temperature minimum observed in the 44$\degree$S-profile near 0.05 mbar ($\sim$320 km) could be due to an oscillation as it is not present in the thermal profile in May 2014 (35$\degree$S). This oscillation is similar to that observed in December 2009 at 74$\degree$N and discussed in Section \ref{discussion: oscillation}. Between May 2014 (35$\degree$S) and October 2014 (46$\degree$S), the temperature decreased between 1 mbar ($\sim$180 km) and 2$\times$10$^{-4}$ mbar ($\sim$575 km), e.g.\ by $\sim$8 K at 0.1 mbar ($\sim$275 km). Then in June 2016 (51$\degree$S), the atmosphere above 0.05 mbar ($\sim$300 km) had warmed up compared to October 2014 (46$\degree$S), by $\sim$11 K at 0.01 mbar ($\sim$375 km), the local temperature maximum, and cooled down below 0.05 mbar by $\sim$6-7 K. Between June 2016 and May 2017 (51$\degree$S), the temperature strongly increased between 0.3 mbar ($\sim$225 km) and 0.01 mbar ($\sim$390 km), by up to $\sim$ 20 K at 0.04 mbar ($\sim$320 km), which is the local temperature maximum. In the lower stratosphere (0.5-5 mbar), the atmosphere cooled by approximately 15 K between 2012 and 2017.

% In May 2014, the stratosphere was well-defined and the stratopause located around $\sim$ 8*10$^{-2}$ mbar as during the northern winter. On the other hand, the stratopause temperature was around $\sim$178 K. Later in October 2014, the temperature profile got colder at all pressure levels. The stratopause was located around 10$^{-1}$ mbar, with a temperature of $\sim$168 K. Then in June 2016, the stratopause wass located higher in altitude at $\sim$8*10$^{-3}$ mbar (value in km) with a temperature of $\sim$176 K. In May 2017, the stratosphere was well-defined and the stratopause is located at $\sim$3.7*10$^{-2}$ mbar with a temperature of $\sim$188 K.

\subsubsection{High southern latitudes}
During southern summer, CIRS did not record limb spectra at 0.5-\cm\ spectral resolution in the south polar region.

Around southern autumn equinox, temperature profiles at high southern latitudes did not present strong variations between August 2009 and January 2010, being actually similar to the thermal profile near the equator in March 2009 below the 0.05-mbar level. Still, we observed a slight cooling between 0.7 mbar ($\sim$195 km) and 0.03 mbar ($\sim$350 km) during this period, reaching $\sim$5 K at 0.1 mbar ($\sim$290 km), while between 0.03 mbar and 0.001 mbar ($\sim$520 km) the temperature increased by $\sim$3 K.

The evolution of the temperature profiles is much more dramatic during southern autumn between 2010 and 2016.  At first, we observe a limited decrease in temperature between January 2010 (74$\degree$S) and July 2010 (80$\degree$S) in the range 0.4-0.05 mbar (220-320 km), reaching $\sim$3 K at 0.2 mbar (255 km), and a slight increase in the range 0.01-0.001 mbar (395-510 km). Fourteen months later, in September 2011 (86$\degree$S), temperature had increased in the range 0.08-4$\times$10$^{-3}$ mbar (280-430 km), by as much as $\sim$7 K at 0.01 mbar (380 km), while the region below the 0.08-mbar level strongly cooled, e.g. by $\sim$16 K at 1 mbar (175 km). In March 2015 (78$\degree$S), the whole region below the 4$\times$10$^{-4}$-mbar (500 km) region had strongly cooled compared to September 2011, by as much as $\sim$26 K at 0.01 mbar ($\sim$350 km). Only six months later, in September 2015 (86$\degree$S), we observe a strong increase in temperature below the 10$^{-4}$-mbar region ($\sim$595 km), reaching $\sim$31 K at 0.04 mbar (285 km). A few months later, in January (86$\degree$S) and February (80$\degree$S) 2016, the thermal profiles had strongly cooled down below 0.01 mbar ($\sim$325 km), by as much as $\sim$30 K at 0.2 mbar (210 km), while they had not changed much
above this pressure level.

%In July 2010, the stratopause was located around  with a temperature around $\sim$ 175 K. Close to the south pole in Septembre 2011, the stratopause was located higher in altitude at around  with a temperature of $\sim$ 176 K. Later in March 2015 at 78$\degree$S, the stratosphere got really cold: the stratopause located higher in altitude at around , and the temperature gradient was shallow between $\sim$ 2*10$^{-1}$ mbar and $\sim$ 3*10$^{-3}$ mbar. From January 2016, the stratosphere was well-defined but the temperature gradient was really steep between 10$^{-2}$ and 1 mbar. The stratopause was located deeper in altitude at around 6*10$^{-2}$ mbar with a temperature of $\sim$ 174 K. In February 2016, the stratopause temperature at 80$\degree$S was higher by $\sim$ 4 K than one month before at 86$\degree$S.

\subsection{Molecular gas volume mixing ratio profiles}
\begin{figure}
\includegraphics[width=\linewidth]{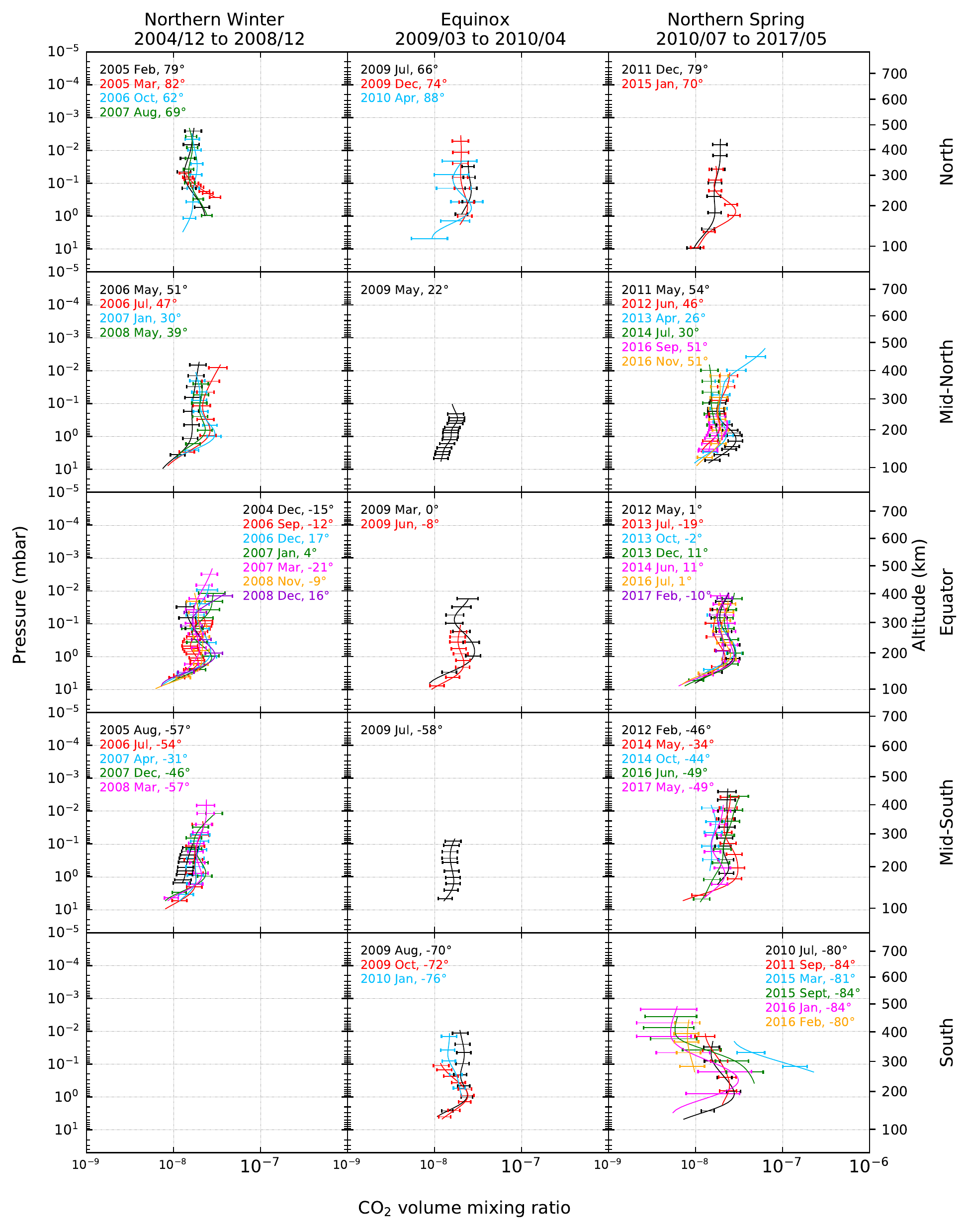}
\caption{\dioxide \/ gas volume mixing ratio profiles in the middle atmosphere of Titan. Same format as Fig.\ \ref{fig:fig_proft_1}.}
\label{fig:fig_profq_co2}
\end{figure}

\begin{figure}
\includegraphics[width=\linewidth]{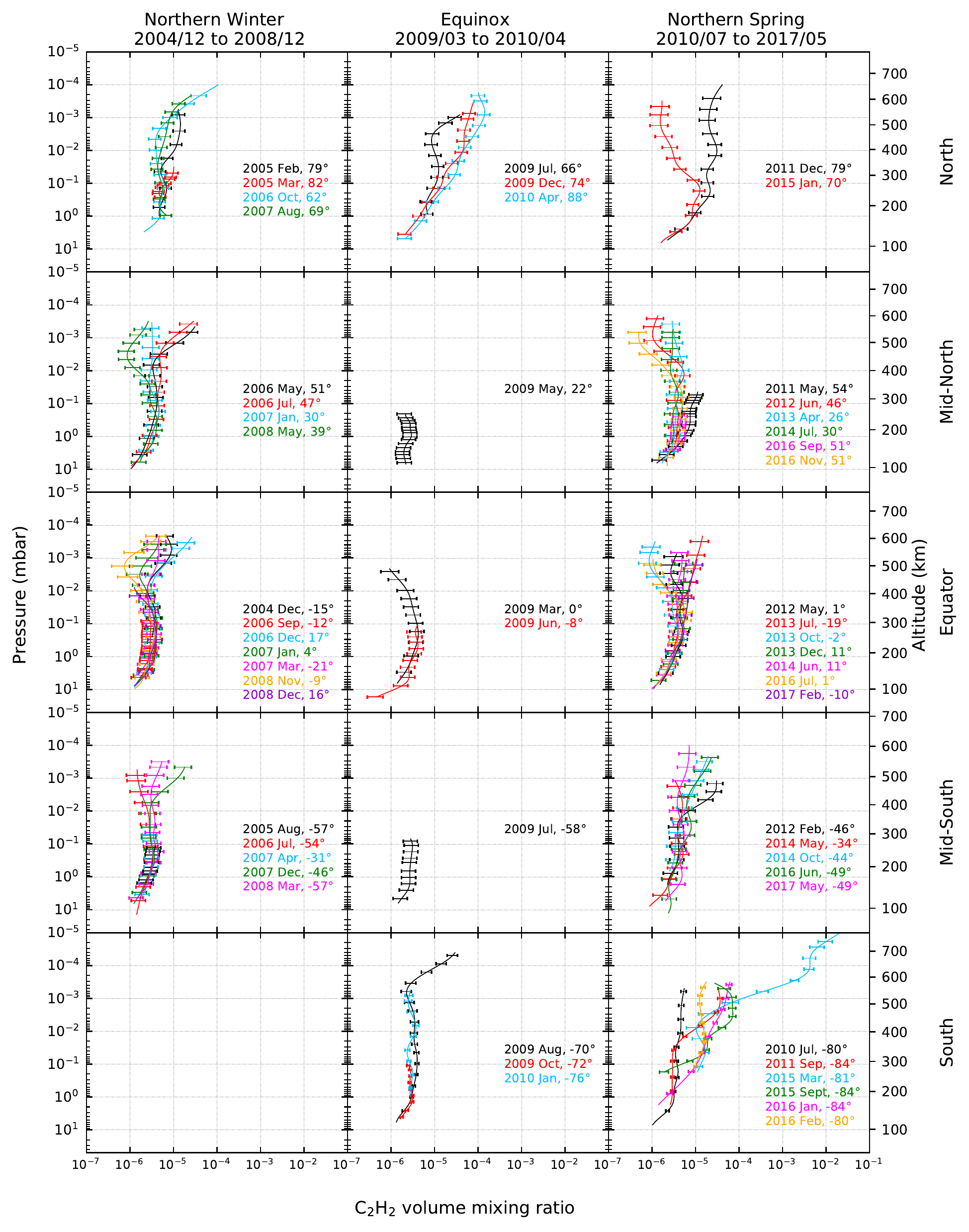}
\caption{\acetylene \/ gas volume mixing ratio profiles in the middle atmosphere of Titan. Same format as Fig.\ \ref{fig:fig_proft_1}.}
\label{fig:fig_profq_c2h2}
\end{figure}

\begin{figure}
\includegraphics[width=\linewidth]{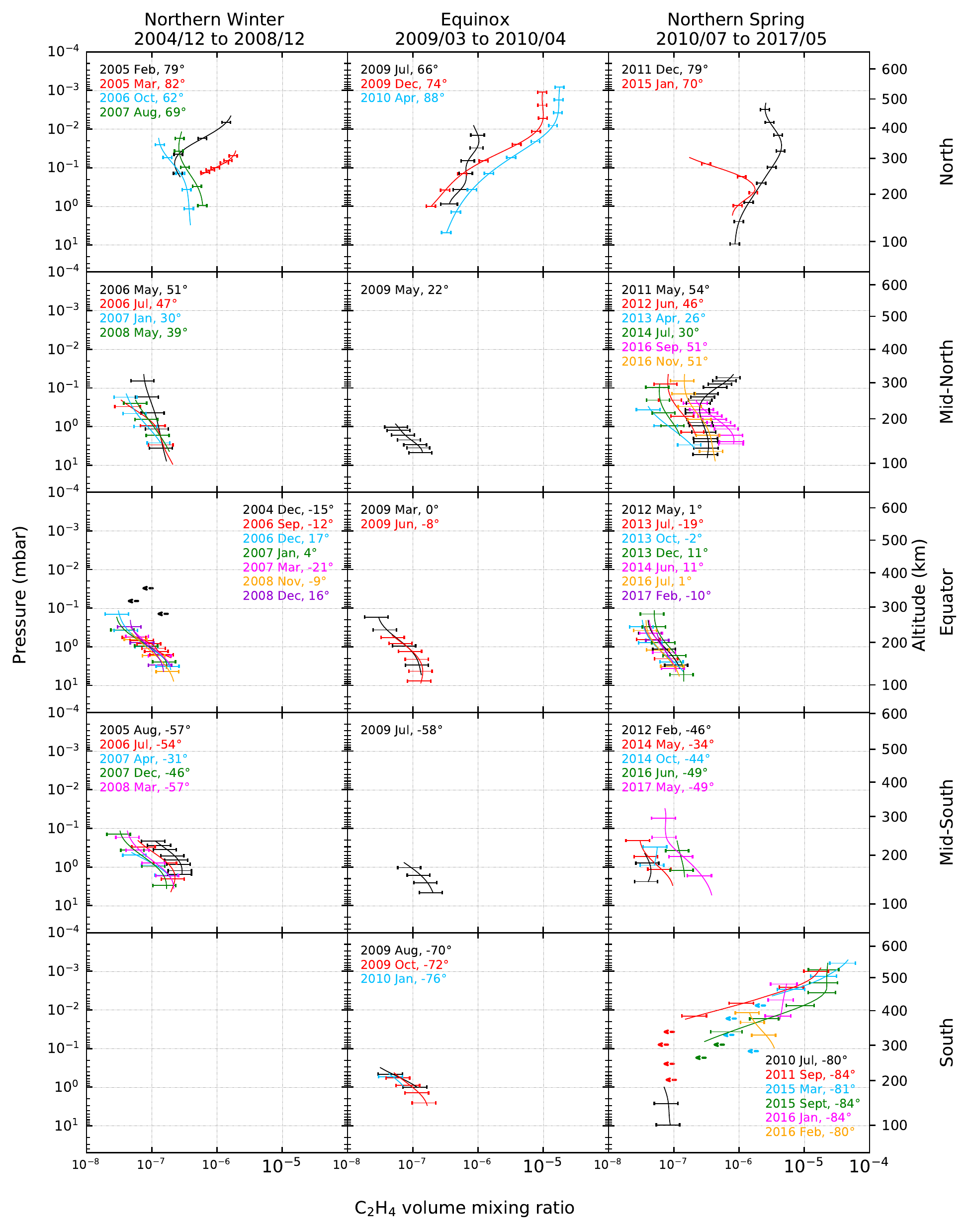}
\caption{\ethylene \/ gas volume mixing ratio profiles in the middle atmosphere of Titan. Same format as Fig.\ \ref{fig:fig_proft_1}.}
\label{fig:fig_profq_c2h4}
\end{figure}

\begin{figure}
\includegraphics[width=\linewidth]{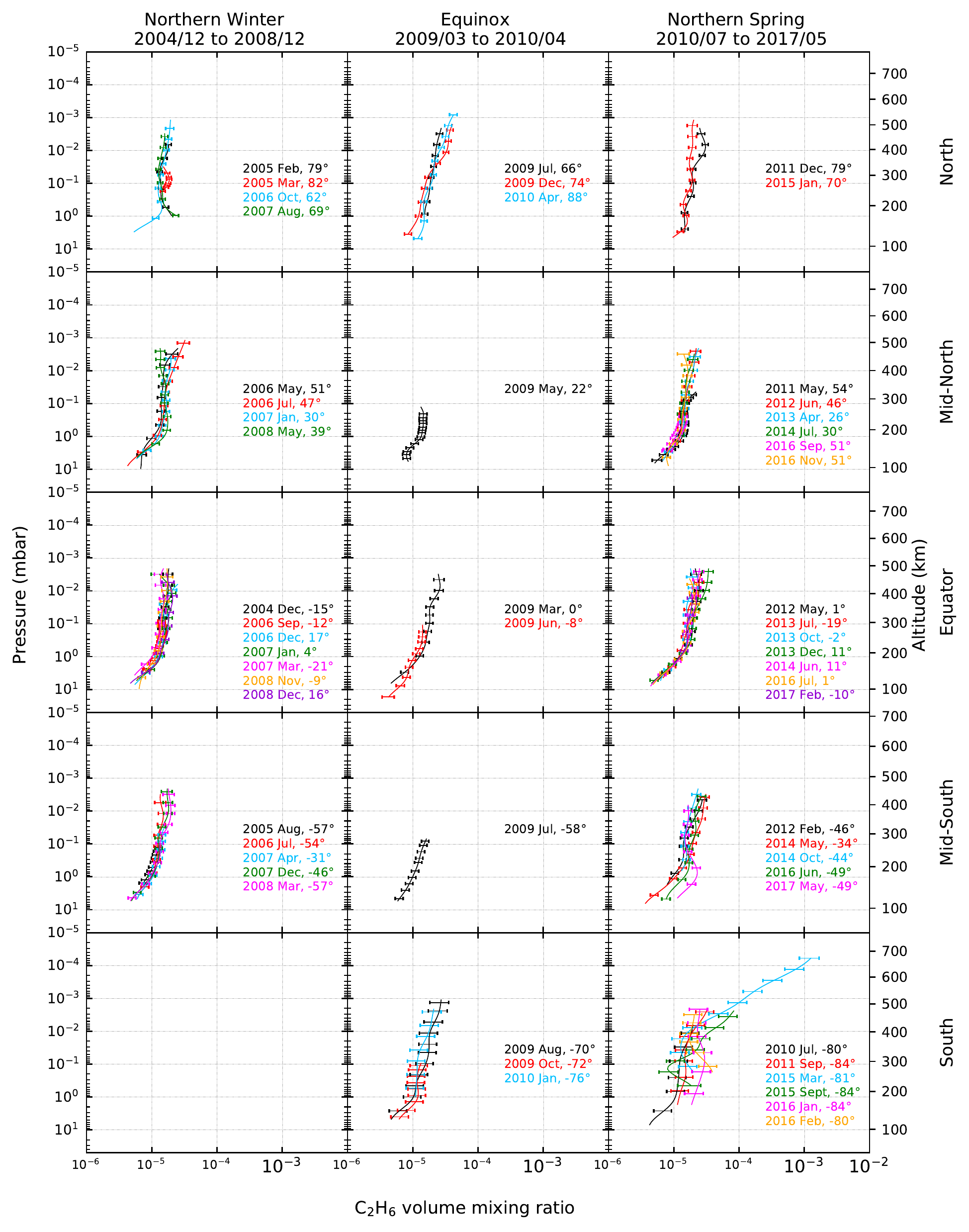}
\caption{\ethane \/ gas volume mixing ratio profiles in the middle atmosphere of Titan. Same format as Fig.\ \ref{fig:fig_proft_1}.}
\label{fig:fig_profq_c2h6}
\end{figure}

\begin{figure}
\includegraphics[width=\linewidth]{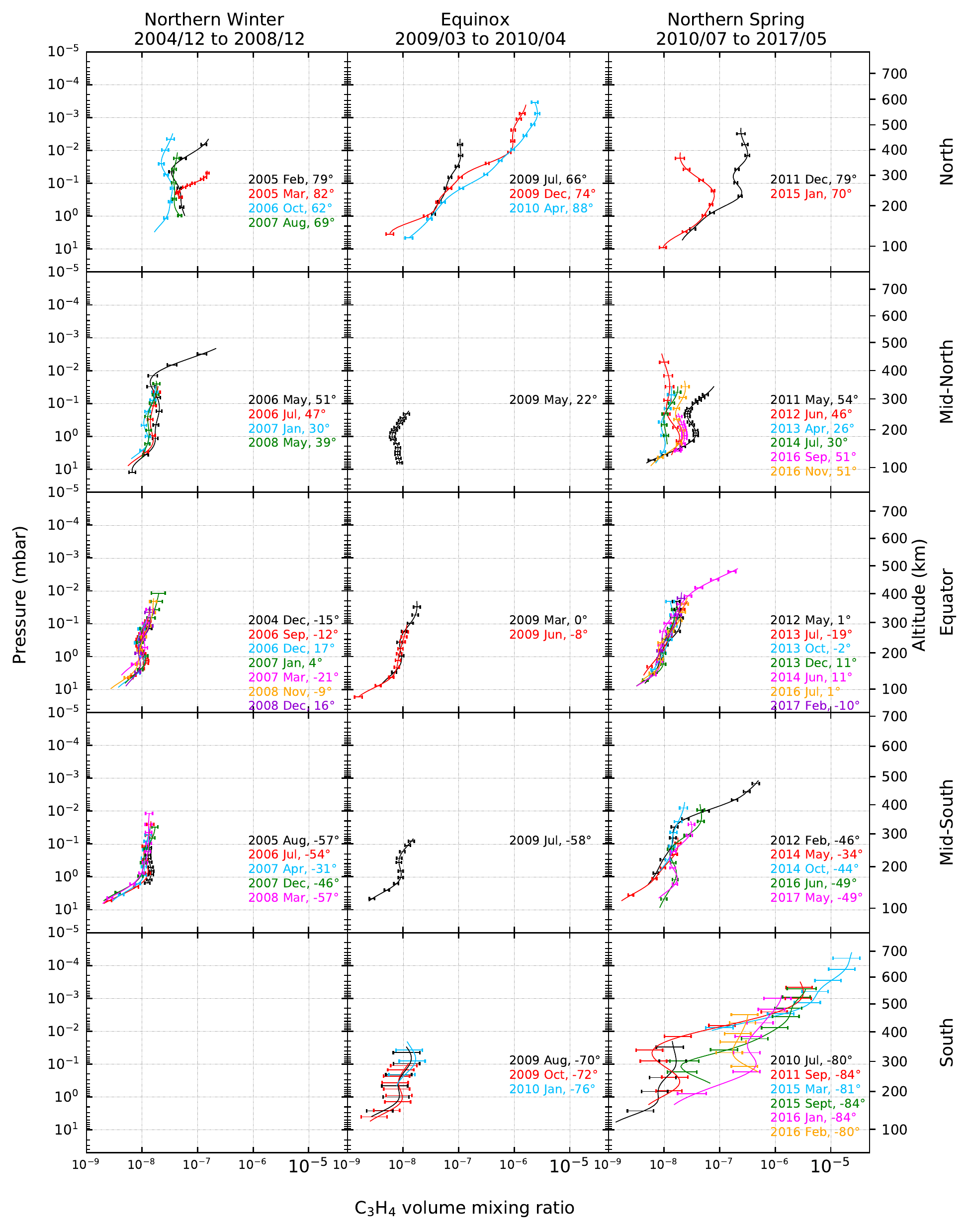}
\caption{\methyl \/ gas volume mixing ratio profiles in the middle atmosphere of Titan. Same format as Fig.\ \ref{fig:fig_proft_1}.}
\label{fig:fig_profq_c3h4}
\end{figure}

\begin{figure}
\includegraphics[width=\linewidth]{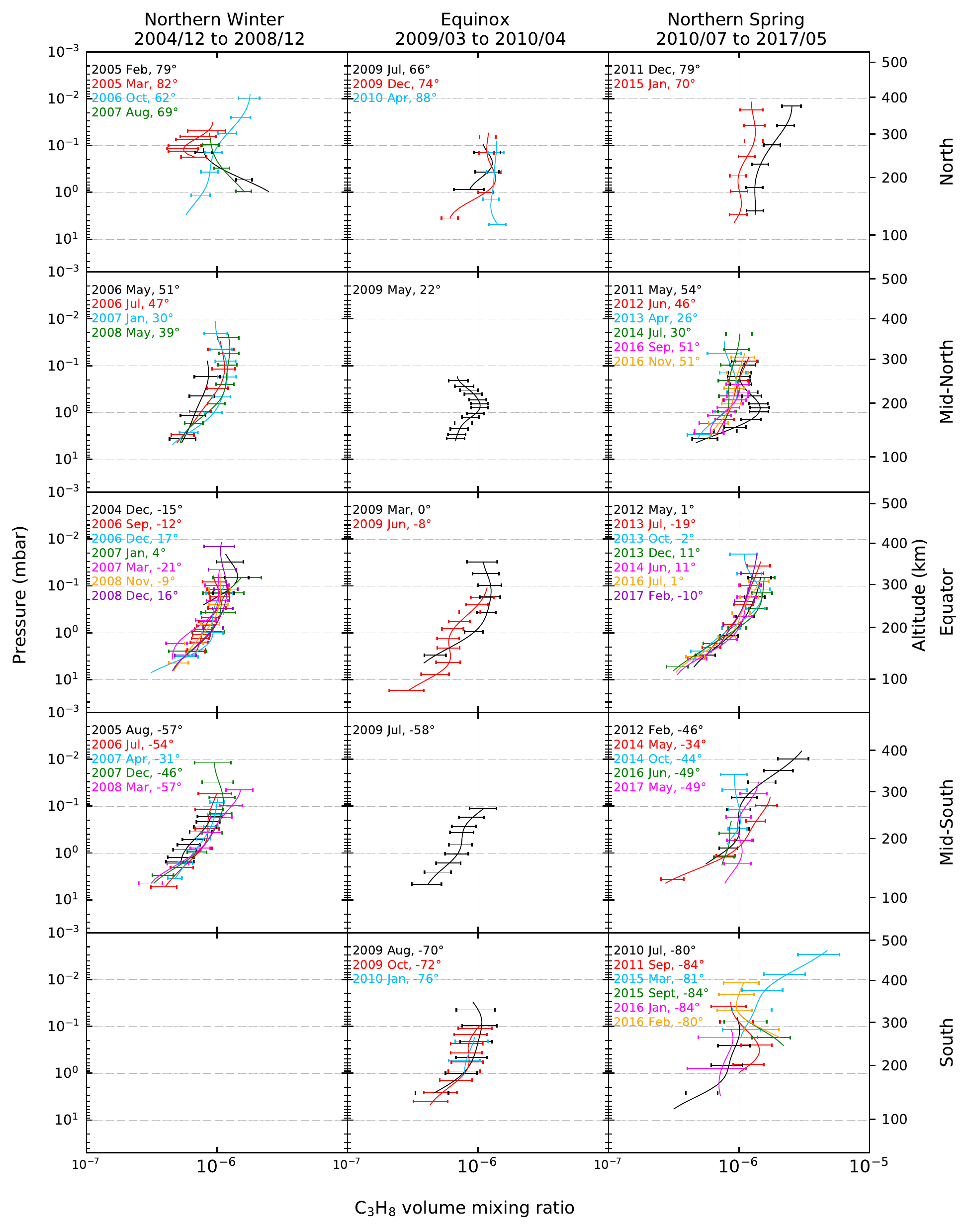}
\caption{\propane \/ gas volume mixing ratio profiles in the middle atmosphere of Titan. Same format as Fig.\ \ref{fig:fig_proft_1}.}
\label{fig:fig_profq_c3h8}
\end{figure}

\begin{figure}
\includegraphics[width=\linewidth]{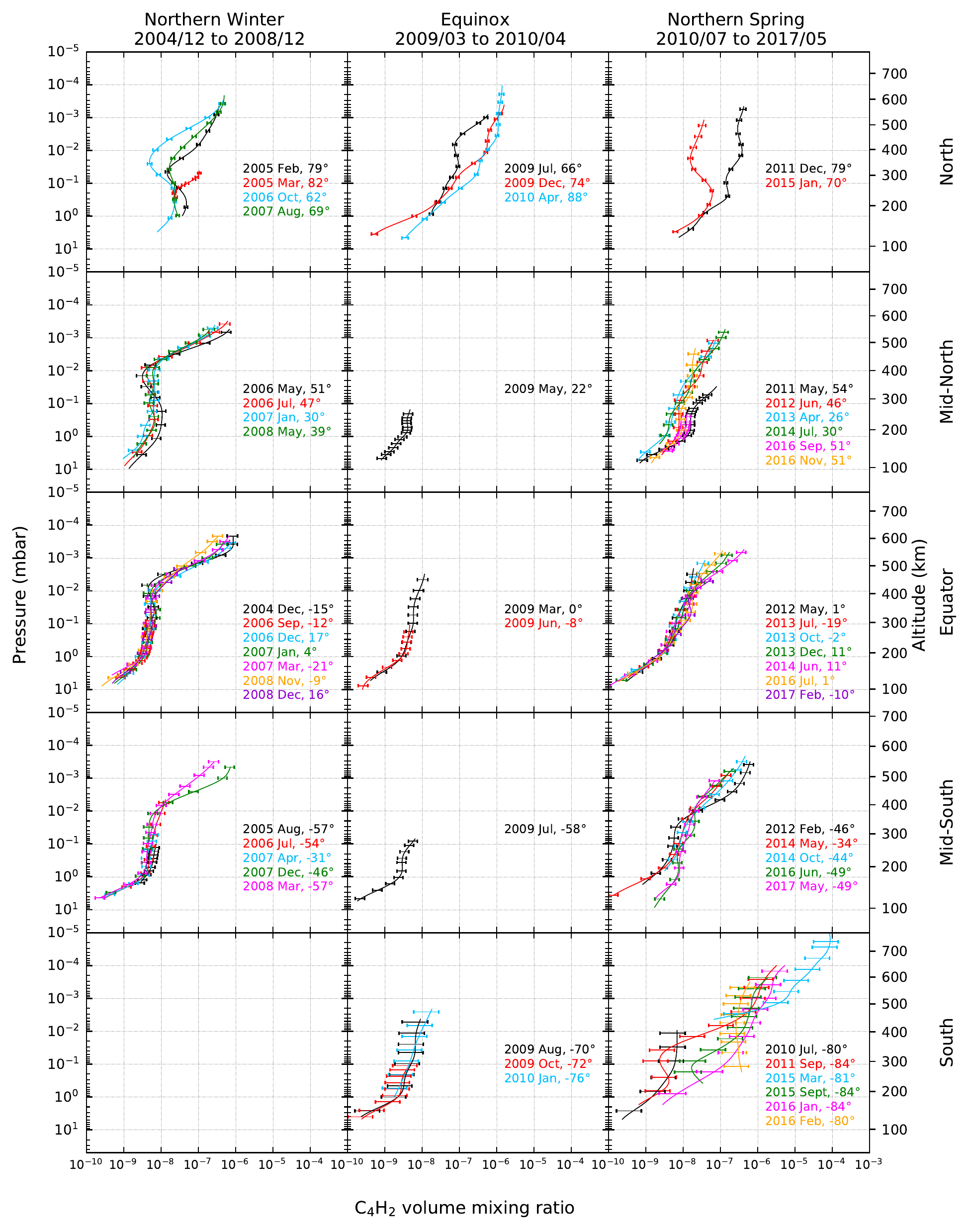}
\caption{\diacetylene \/ gas volume mixing ratio profiles in the middle atmosphere of Titan. Same format as Fig.\ \ref{fig:fig_proft_1}.}
\label{fig:fig_profq_c4h2}
\end{figure}

\begin{figure}
\includegraphics[width=\linewidth]{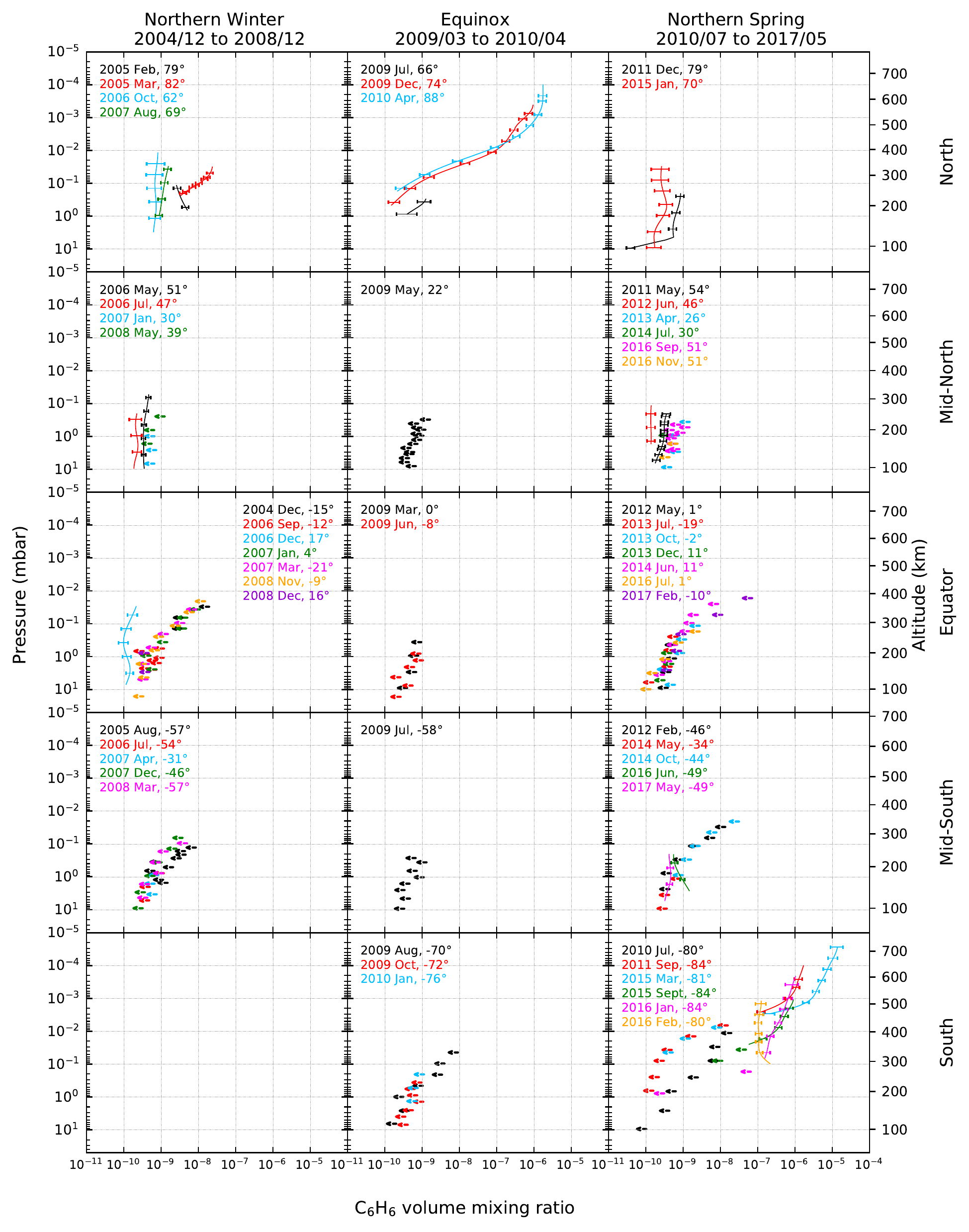}
\caption{\benzene \/ gas volume mixing ratio profiles in the middle atmosphere of Titan. Same format as Fig.\ \ref{fig:fig_proft_1}.}
\label{fig:fig_profq_c6h6}
\end{figure}

\begin{figure}
\includegraphics[width=\linewidth]{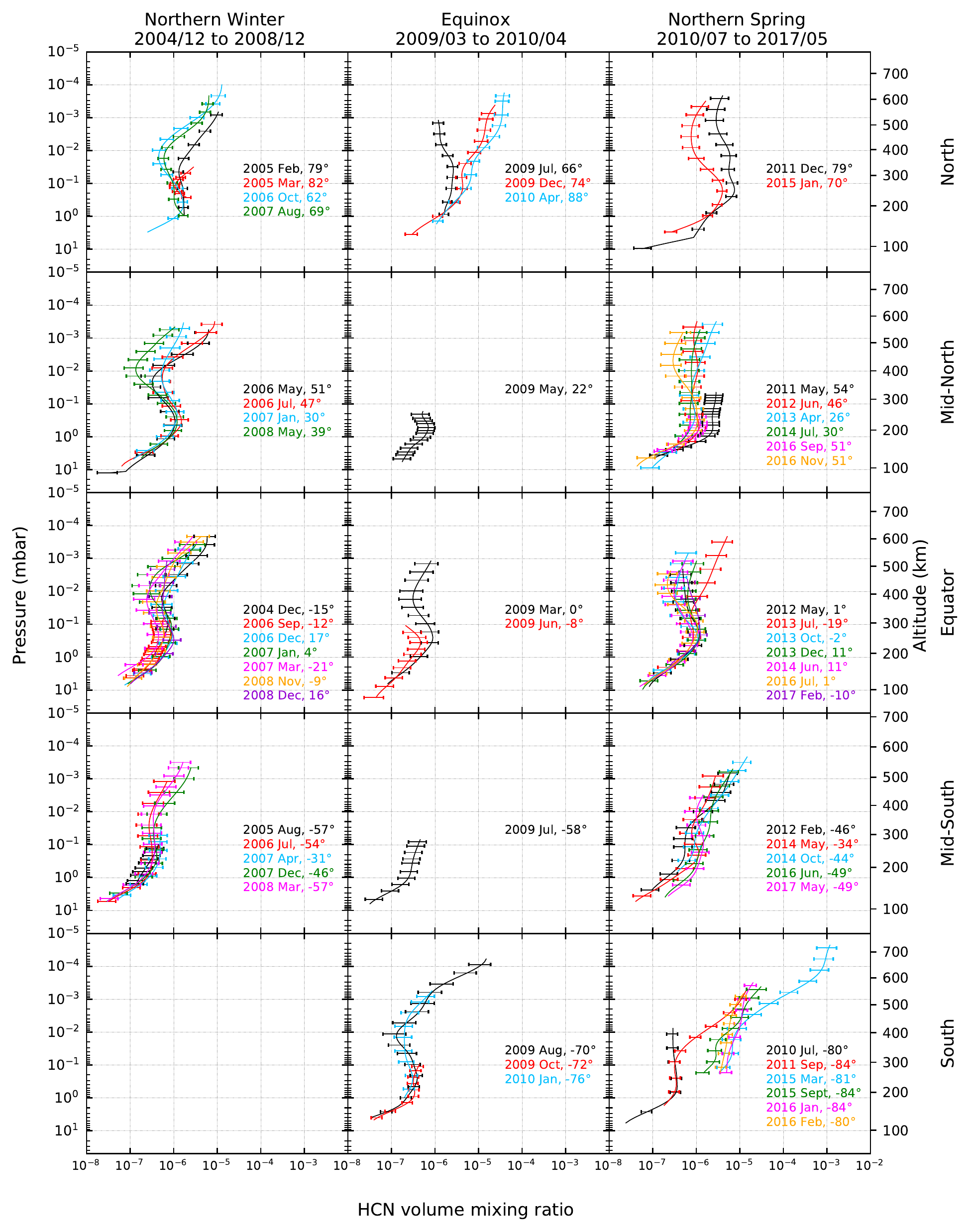}
\caption{HCN gas volume mixing ratio profiles in the middle atmosphere of Titan. Same format as Fig.\ref{fig:fig_proft_1}.}
\label{fig:fig_profq_hcn}
\end{figure}

\begin{figure}
\includegraphics[width=\linewidth]{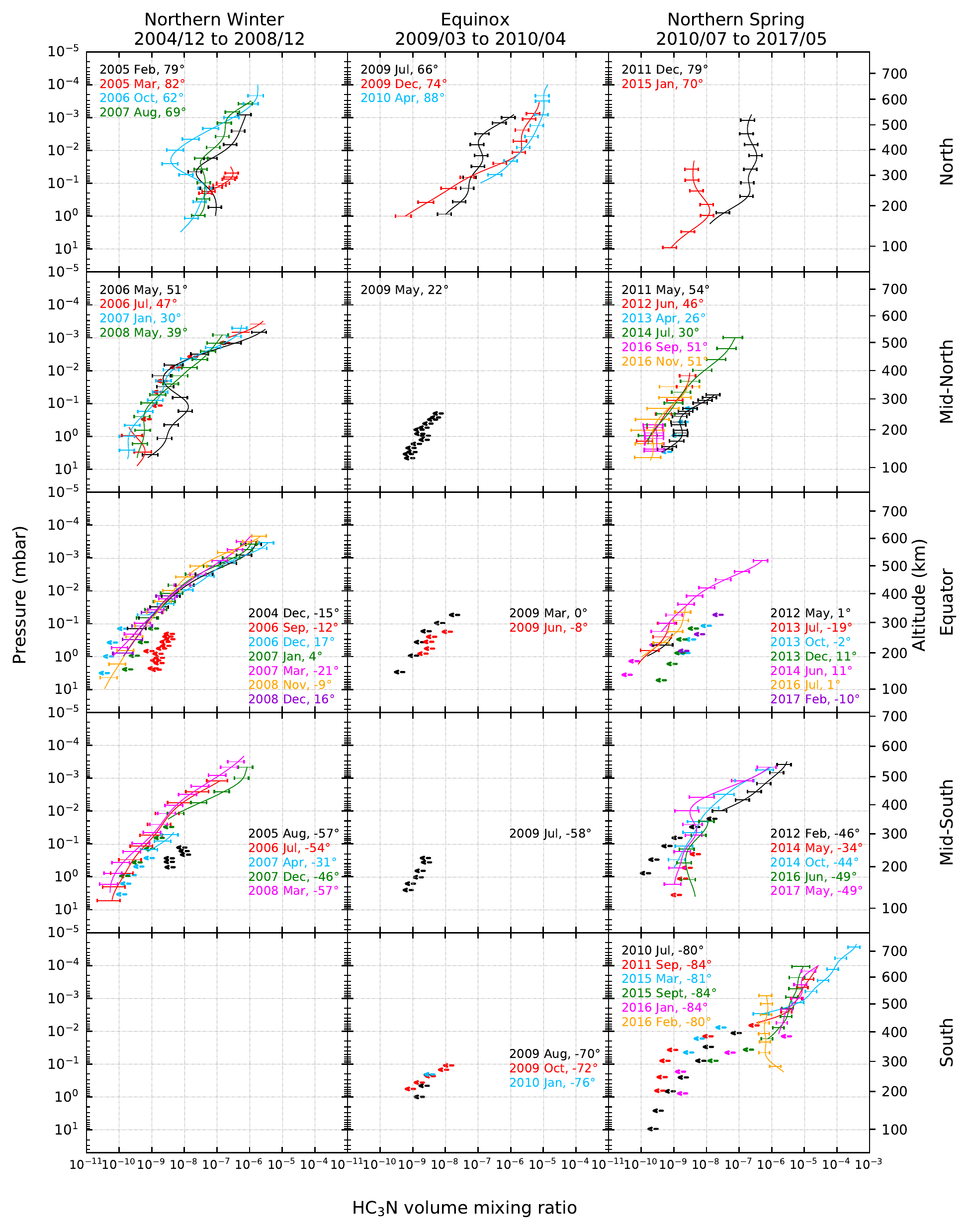}
\caption{\cyano \/ gas volume mixing ratio profiles in the middle atmosphere of Titan.  Same format as Fig.\ \ref{fig:fig_proft_1}.}
\label{fig:fig_profq_hc3n}
\end{figure}

Figures \ref{fig:fig_profq_co2} to \ref{fig:fig_profq_hc3n} show the gas mixing ratio profiles of: C$_2$H$_2$, C$_2$H$_4$, C$_2$H$_6$, C$_3$H$_4$, C$_3$H$_8$, C$_4$H$_2$, C$_6$H$_6$, HCN, HC$_3$N and CO$_2$ retrieved using the limb observations listed in Table \ref{table:list_obs}.

Carbon dioxide (CO$_2$) does not exhibit strong seasonal variations at any latitude. The profiles show a local maximum around 1 mbar ($\sim$190 km) at all latitudes, except during the northern winter at high northern latitudes (Fig.\ \ref{fig:fig_profq_co2}). The peak-to-peak amplitude of this variation is at most a factor of two. We will not discuss further the CO$_2$ mixing ratio profiles in the following sections.

\subsubsection{High northern latitudes}
\begin{figure}
\includegraphics[width=\linewidth]{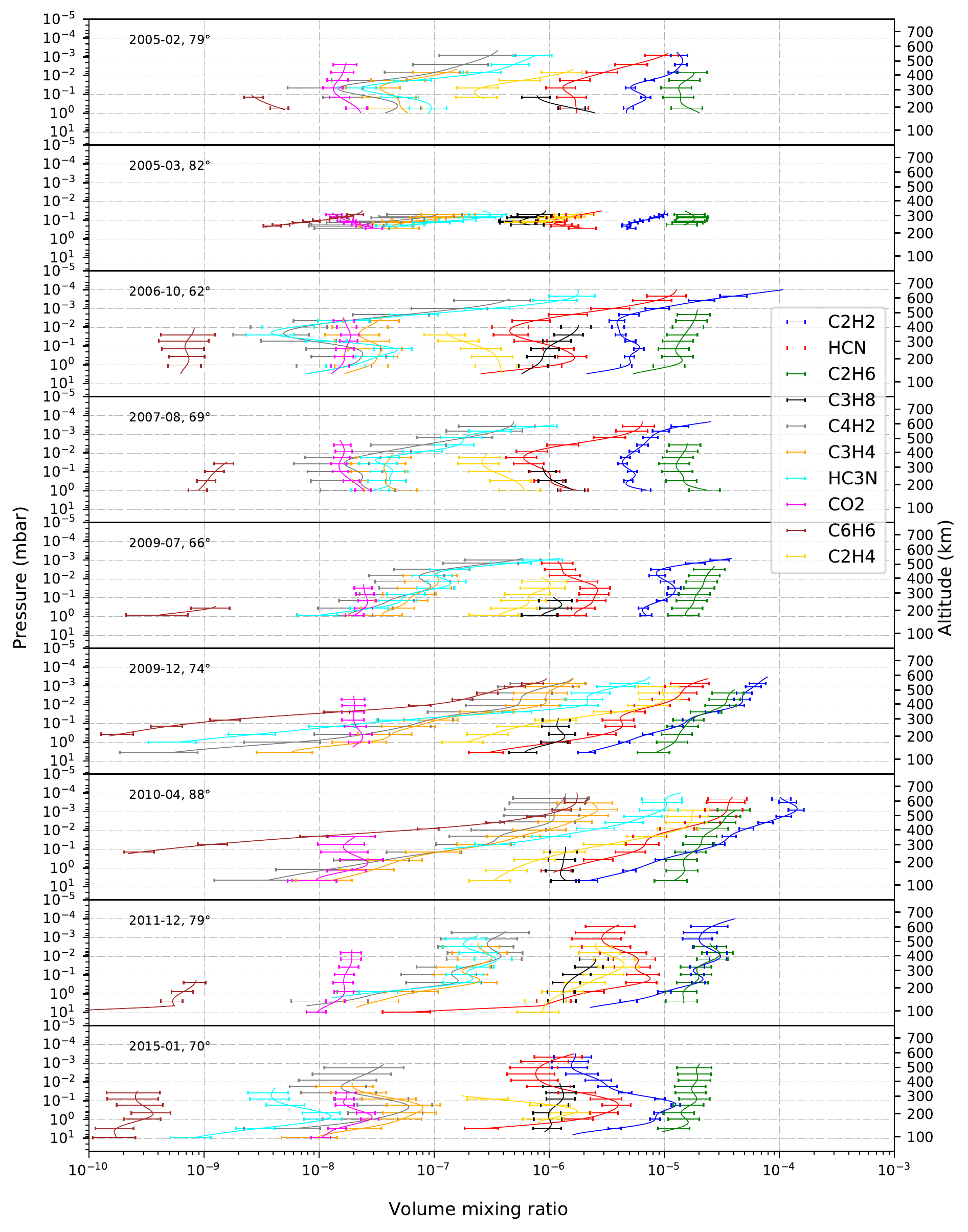}
\caption{Gas volume mixing ratio profiles from February 2005 to January 2015 in the northern polar region. The altitude scale corresponds to the October 2006 observations.}
\label{fig:north_profqs}
\end{figure}

Seasonal changes of abundance profiles in the northern polar region are also shown in Fig.\  \ref{fig:north_profqs}. During northern winter, from February 2005 to August 2007, the abundances of most photochemical compounds between 10 mbar ($\sim$100 km) and 10$^{-4}$ mbar ($\sim$690 km) were larger than at equatorial latitudes at the same season. The difference amounts to a factor of 3-5 for C$_3$H$_4$, C$_4$H$_2$, C$_2$H$_4$, C$_6$H$_6$ and a factor of 50 for HC$_3$N at 1 mbar ($\sim$175 km). On the other hand, the C$_2$H$_6$ and C$_3$H$_8$ profiles did not show significant differences with their equatorial counterparts. During this period, we also detected benzene below 0.01 mbar ($\sim$400 km) with a mole fraction of $\sim$4$\times$10$^{-9}$ at 80$\degree$N, 10$^{-9}$ at 70$\degree$N, and 3-5$\times$10$^{-10}$ at 50$\degree$N.

\begin{figure}
\includegraphics[width=\linewidth]{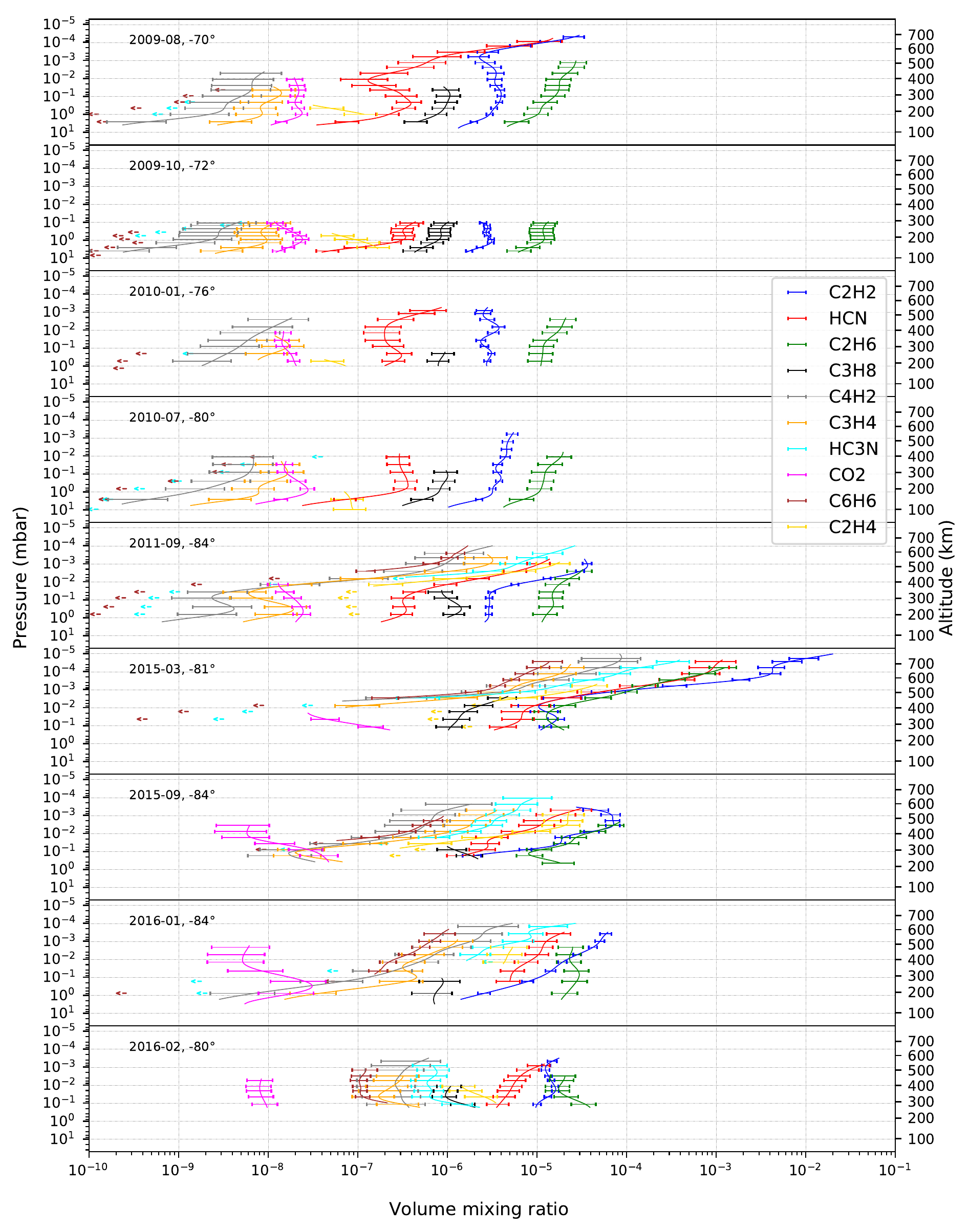}
\caption{Gas volume mixing ratio profiles from August 2009 to February 2016 in the southern polar region. The altitude scale corresponds to the October 2009 observations.}
\label{fig:south_profqs}
\end{figure}

In December 2009 and April 2010, shortly after northern spring equinox, we observed an enrichment in photochemical compounds above 0.1 mbar ($\sim$260 km), compared to July 2009. In particular, benzene was detected up to the 10$^{-4}$-mbar level ($\sim$630 km). We can also note that C$_2$H$_2$ was more abundant than C$_2$H$_6$ at pressures less than a few 10$^{-2}$ mbar ($\sim$375 km) (Fig.\ \ref{fig:north_profqs}).

In the beginning of northern spring in December 2011, we observed a global depletion in molecules, except for C$_2$H$_6$, above 0.1 mbar ($\sim$260 km) with respect to the April 2010 profiles. We also observed that the C$_2$H$_2$ abundance profile was similar to the C$_2$H$_6$ profile, and that benzene was only observed below the 0.01-mbar ($\sim$375 km) level with a mixing ratio of a few 10$^{-10}$ at 79$\degree$N. Then in January 2015, the depletion in molecules above 0.1 mbar ($\sim$280 km) was even more pronounced than in December 2011, by factors of $\sim$12,  $\sim$8 and $\sim$3 for C$_2$H$_2$,  C$_4$H$_2$ and HCN respectively, at 1$\times$10$^{-3}$ mbar ($\sim$480 km).

\subsubsection{Mid and equatorial latitudes}
Between 60$\degree$N and 60$\degree$S in the stratosphere below 0.01 mbar ($\sim$410 km), photochemical compounds are slighly more abundant during northern winter season than in northern spring. Above 0.01 mbar, C$_2$H$_2$ and HCN abundance profiles show variations that do not seem correlated with year or local time (see Table \ref{table:list_obs}).

However, we observe in some specific cases abundance variations at latitudes close to the winter polar vortex. In May 2006 (51$\degree$N), a local maximum of the HC$_3$N mixing ratio occurs at 0.1 mbar ($\sim$290 km), as is the case at 62$\degree$N in October 2006. Around 45$\degree$S, we observed an enrichment in all photochemical compounds below 0.1 mbar ($\sim$290 km) between October 2014 and June 2016, as is seen at higher southern latitudes (Section \ref{High southern latitudes}).

\subsubsection{High southern latitudes}
\label{High southern latitudes}
Seasonal changes of abundance profiles in the southern polar region are also shown in Fig.\ \ref{fig:south_profqs}. From August 2009 (70$\degree$S) to at least July 2010 (80$\degree$S), we did not observe strong variations in any abundance profile between 10 mbar ($\sim$95 km) and 10$^{-3}$ mbar ($\sim$520 km). In September 2011, during southern autumn, we observed a strong and global enrichment in photochemical products compared to June 2010 between 10$^{-2}$ mbar ($\sim$380 km) and 10$^{-4}$ mbar ($\sim$620 km). This enrichment reaches a factor of 100-1000 for molecules such as C$_3$H$_4$, C$_4$H$_2$, HC$_3$N, C$_2$H$_4$ and a factor of 10 for molecules such as C$_2$H$_2$, HCN. At this date, we also detected benzene in the southern polar region above 2$\times$10$^{-3}$ mbar ($\sim$465 km) with a mixing ratio of at least 10$^{-7}$-10$^{-6}$. Then by March 2015, molecular abundances had continued to increase, being larger than in September 2011 by a factor of 10 above 10$^{-3}$ mbar (455 km). Between 0.1 mbar (255 km) and 10$^{-3}$ mbar (455 km), C$_2$H$_2$ and HCN mixing ratios were larger by factors of 3 and 10 respectively. In September 2015, only six months later, the abundances of the photochemical compounds   had significantly decreased near and above 10$^{-3}$ mbar ($\sim$475 km) but were still larger than inferred around southern autumn equinox. This decrease continued at a lower pace in the 10$^{-3}$-mbar region till January 2016. One month later, the abundance profiles we inferred were relatively constant with height between 0.1 mbar ($\sim$230 km) and 10$^{-3}$ mbar ($\sim$450 km), except for HCN which showed a factor of 3 variation over this altitude range.
 % We noticed C$_2$H$_6$ and C$_3$H$_8$ are the least impacted by the atmospheric dynamics of the global circulation, probably because of their long chemical lifetimes \citep{Vuitton2019}.

%============================================================================%
% PART: DISCUSSION
%============================================================================%
\section{Discussion}

\subsection{Equatorial mixing ratio profiles compared with photochemical models}
%%%
% comparer avec les modèles
% pole: GCM pour la dynamique des pôles, Lebonnois 2012, Newman 2016, Mitchell 2016
% equator : photochimie Krasnopolsky2014, Loison2015, Dobrijevic2016, Vuitton 2018
%%%
% comparer avec les autres observations
% CIRS : Vinatier2015, Coustenis2018, Sylvestre2018 (15mbar)
% UVIS : Koskinen2011
% ALMA : Cordiner2014, Thelen2019
%%%
%As the equatorial regions do not vary strongly, we can compared with photochemical model and local maximal volume mixing ratio value. 
We have shown that temperature and volume mixing ratio profiles do not change significantly over seasons at the equator from 10 ($\sim$90 km) to 0.01 mbar ($\sim$290 km), which is consistent with previous CIRS investigations \citep{Achterberg2008, Vinatier2015, Sylvestre2018, Teanby2019}. On the other hand, we found that the C$_2$H$_2$ (acetylene) mole fraction varies in the mesosphere above 10$^{-2}$ mbar ($\sim$400 km) without any correlation with the local time (see Table \ref{table:list_obs}). Such variations are not currently explained. 

%In fact, the abundance profiles of molecules at the equator slightly increase between 10 mbar ($\sim$ 90 km) and 0.1 mbar ($\sim$ 290 km) from the northern winter to the northern spring.

% local maximal et minimal value mixing ratio
%A local maximal value of volume mixing ratio is shown for species with molecules with the longest lifetime such as C$_2$H$_2$, HCN, C$_3$H$_8$ near the stratopause level at 10$^{-1}$ mbar. This is not consistent with photochemical models \citep{Krasnopolsky2014, Loison2015, Dobrijevic2016, Vuitton2019}. These studies show a constant volume mixing ratio profile above 200 km for C$_3$H$_8$, slightly increasing with height above 200 km for C$_2$H$_2$ and HCN \citep{Loison2015, Dobrijevic2016, Vuitton2019}. 

%A local minimal value of the volume mixing ratio is found for C$_2$H$_2$, HCN and C$_4$H$_2$ abundances profiles at around 10$^{-2}$ mbar during the northern winter. This local minimum seems to disappear during the northern spring equinox transition, except for HCN which still shows a maximum in March 2009 at 0$\degree$N. In contrast, in the northern spring, a shallow local minimum is observed for HCN at 3*10$^{-2}$ mbar (value in km) until May 2014 (34$\degree$S), moving upward to 10$^{-2}$ mbar (value in km) in May 2017 (49$\degree$S).

% C2H2
If we focus on the 10-0.01 mbar region, our derived C$_2$H$_2$ abundance profiles show a local maximum at $\sim$0.2 mbar ($\sim$265 km) that it is not predicted by photochemical models \citep{Krasnopolsky2014,Loison2015,Dobrijevic2016,Vuitton2019}. At 0.2 mbar, the observed C$_2$H$_2$ volume mixing ratio of $\sim$4$\times$10$^{-6}$ is consistent within a factor of 2 with photochemical model predictions. The C$_2$H$_2$ abundance derived in November 2008 at 5$\times$10$^{-4}$ mbar ($\sim$500 km) is consistent with the Cassini/UVIS C$_2$H$_2$ abundance derived in February 2008 at 550 km at 6$\degree$S \citep{Koskinen2011}. 
Above 550 km, our retrieved C$_2$H$_2$ profiles are quite consistent with C$_2$H$_2$ profiles derived from VIMS spectra \citep{Dinelli2019}.

% HCN
Our inferred HCN (hydrogen cyanide) equatorial abundance profiles show a local maximum around 0.3 mbar ($\sim$245 km) and a local minimum around 0.01 mbar ($\sim$410 km), both of which are not reproduced by photochemical models \citep{Loison2015,Krasnopolsky2014,Dobrijevic2016,Vuitton2019}, while at 0.3 mbar, the derived mixing ratio of $\sim$10$^{-6}$ agrees with these model predictions. We note that the eddy mixing coefficient plays an important role in the HCN profile as simulated by \cite{Vuitton2019} who obtained the best match with our retrieved profiles using an eddy mixing coefficient having  a minimum value $K_0$ $\sim$ 100 cm$^2$ s$^{-1}$, instead of $K_0$~= 300 cm$^2$ s$^{-1}$ in the nominal model. In November 2008, our HCN abundance at 500 km ($\sim$7$\times$10$^{-6}$) is consistent with the mixing ratio derived from UVIS stellar occultation measurements at 550 km in February 2008 at 6$\degree$S \citep{Koskinen2011}. Above 550 km, our retrieved HCN profiles are less abundant than those derived from VIMS spectra \citep{Dinelli2019}.

%Profiles retrieved in this study reach altitudes ($\gtrsim$ 550km) where the non-LTE effect is no more negligible. \cite{Dinelli2019} analyzed VIMS spectra in the upper atmosphere using a non-LTE model to retrieve CH$_4$, C$_2$H$_2$ and HCN abundance profiles. Our C$_2$H$_2$ profiles are quite consistent with their C$_2$H$_2$ profiles, but our HCN profiles are always less abundant than their HCN profiles which the gap is more marked during the northern spring.}

% C2H4
The C$_2$H$_4$ (ethylene) mixing ratios we inferred around 0.1 mbar ($\sim$290 km) are consistent with predictions of the photochemical models of \cite{Krasnopolsky2014}, \cite{Loison2015} and \cite{Dobrijevic2016}, but  less abundant by a factor of 5 than that of \cite{Vuitton2019}. From 5 to 0.1 mbar, the negative vertical gradient we derived is at odds with these photochemical models. According to \cite{Crespin2008}, it is explained by the fact that C$_2$H$_4$ does not condense in Titan's conditions and by dynamical advection that transports air enriched in C$_2$H$_4$ from the winter pole towards the equator in the lower stratosphere.

% C3H8
For C$_3$H$_8$ (propane), the inferred abundance profiles at the equator are consistent in the range 10-0.01 mbar ($\sim$90-410 km) with the photochemical model predictions of \cite{Krasnopolsky2014}, \cite{Loison2015}, \cite{Dobrijevic2016} and \cite{Vuitton2019}.

% C3H4
Regarding C$_3$H$_4$ (methylacetylene), our abundance profiles show a mixing ratio almost constant with height between 3 mbar ($\sim$140 km) and 0.1 mbar ($\sim$300 km), which is consistent with \cites{Vuitton2019} and \cites{Krasnopolsky2014} photochemical models, while in this altitude range, \cites{Loison2015} and \cites{Dobrijevic2016} photochemical models predict a C$_3$H$_4$ abundance decreasing with height. At 0.02 mbar, our C$_3$H$_4$ mixing ratios in January 2007 and after 2012 are consistent with these photochemical models. in June 2014 above 0.01 mbar ($\sim$410 km), we derived a positive abundance gradient consistent with \cites{Vuitton2019} photochemical model, while this change of slope is observed at 600 km in \cites{Loison2015} and \cites{Dobrijevic2016} models and at 300 km in \cites{Krasnopolsky2014} photochemical model.

% C4H2
% UVIS : 5e-7 à 500 km vers 6°S
% Vuitton : plutot la courbe avec no H heterogeneous loss reaction, minimum local vers 400 km par forcément observé mais qui correspond a notre changement de pente
% Krasno : pas du tout car son profil est fortement croissant entre 200 et 400 km
% Loison : a pas C4H2
The C$_4$H$_2$ (diacetylene) abundance profiles retrieved during the northern winter between 0.8 mbar ($\sim$200 km) and 0.01 mbar ($\sim$410 km) are in agreement with the predicted abundance profile of \cites{Dobrijevic2016} photochemical model, and with the \cite{Vuitton2019} photochemical model with no H heterogeneous loss reaction. \cites{Krasnopolsky2014} photochemical model predicts a C$_4$H$_2$ abundance profile with a strong positive gradient in this altitude region that we do not observe. \cites{Dobrijevic2016} and \cites{Vuitton2019} photochemical models predict a local mixing ratio minimum around 500 km that is not observed in our C$_4$H$_2$ abundance profiles at any season. On the other hand, these photochemical models exhibit a positive abundance gradient above 500 km that we observe above 0.01 mbar ($\sim$410 km) during the northern winter. At higher altitudes, our C$_4$H$_2$ mixing ratio of $\sim$7$\times$10$^{-7}$ at 5$\times$10$^{-4}$ mbar (560 km) during the northern winter is consistent with the derived UVIS value of 5$\times$10$^{-7}$ around 500 km in February 2008 at 6$\degree$S \citep{Koskinen2011}. During the northern spring, our C$_4$H$_2$ profiles show a positive abundance gradient between 10 mbar ($\sim$90 km) and 10$^{-3}$ mbar ($\sim$530 km) that is not as strong as the gradient in \cites{Krasnopolsky2014} photochemical model, and is not predicted by \cite{Dobrijevic2016} and \cite{Vuitton2019}.

% HC3N
We were able to detect HC$_3$N (cyanoacetylene) emission at altitudes up to 590 km (2-3$\times$10$^{-4}$ mbar) during the northern winter, which is higher than what was achieved in a previous study at the equator \citep{Vinatier2015}. The present detection of HC$_3$N at high altitude results from the improvement of the signal-to-noise ratio in limb spectra due to the upgraded calibration and our averaging procedure described in Appendix B. HC$_3$N reaches a volume mixing ratio of 2$\times$10$^{-6}$ at $\sim$3$\times$10$^{-4}$ mbar ($\sim$580 km), quite consistent with the $\sim$10$^{-6}$ value derived at 550 km in February 2008 at 6$\degree$S from UVIS stellar occultations \citep{Koskinen2011}. At 3$\times$10$^{-4}$ mbar, our HC$_3$N mixing ratio values are $\sim$14 times larger than the predicted value of \cite{Vuitton2019} when they use the aerosol optical depth from \cite{Lavvas2010} multiplied by a factor of 2. Indeed, HC$_3$N is strongly sensitive to UV photodissociation, so the higher the aerosol opacity the lower the photodissociation efficiency of HC$_3$N because of aerosol UV-shielding. In \cites{Loison2015} and \cites{Dobrijevic2016} models, which incorporate a different aerosol optical depth profile, the HC$_3$N mixing ratios are $\sim$300 times smaller than our value at 3$\times$ 10$^{-4}$ mbar. In contrast, our HC$_3$N mixing ratios are a factor of 30 smaller than in \cites{Krasnopolsky2014} photochemical model, in which the aerosol optical depth model derives from Huygens observations. During the northern winter, we did not find the local minimum in the mixing ratio predicted at 500 km by photochemical models \citep{Vuitton2019, Loison2015, Dobrijevic2016, Krasnopolsky2014}.

% CO2
Regarding CO$_2$ (carbon dioxide), our abundance profiles are consistent with \cites{Krasnopolsky2012} photochemical model, but are higher than the \cite{Loison2015} and \cite{Vuitton2019} profiles  by an order of magnitude. Interestingly, we observe a local maximum at 1 mbar ($\sim$190 km) and a local minimum at 0.1 mbar ($\sim$300 km), which are not predicted by photochemical models \citep{Krasnopolsky2012, Loison2015, Vuitton2019}. This local maximum is also observed at mid-latitudes and at the south pole and seems to be stable with season.
 
% C2H6
Our inferred C$_2$H$_6$ (ethane) abundance profiles are quite consistent with the \cite{Vuitton2019} photochemical model when they use a minimum value of the eddy mixing coefficient somewhat lower than the nominal value of 300 cm$^2$ s$^{-1}$ or when they use half of the aerosol optical depth from \cite{Lavvas2010}. As for HC$_3$N, UV aerosol shielding limits the efficiency of the photodissociation. In \cites{Vuitton2019} model, the C$_2$H$_6$ abundance is very sensitive to the aerosol optical depth, since for a nominal aerosol optical depth C$_2$H$_2$ is less photodissociated, which leads to less C$_2$H and CH$_3$ radicals, the precursors of C$_2$H$_6$. The C$_2$H$_6$ profiles predicted by the photochemical models of \cite{Loison2015} and \cite{Dobrijevic2016}, who used different aerosol depth profile and CH$_3$ recombination rate, are consistent with our retrieved profile. \cites{Krasnopolsky2014} photochemical model, with an aerosol optical depth model based on Huygens observations and a different CH$_3$ recombination rate, has half less ethane than observed between 100 and 250 km, and above this altitude range, the predicted strong abundance gradient is not consistent with our C$_2$H$_6$ abundance profiles.

% C6H6 : il y a que Vuitton et Krasno
Regarding \benzene \/ (benzene), we could only obtain 2-$\sigma$ upper limits of its mixing ratio between 20 mbar ($\sim$80 km) and 0.01 mbar ($\sim$410 km), except for December 2006 where we could detect its emission band. According to photochemical models, the low abundance of \benzene\/ below the production region results from efficient photodissociation to phenyl radicals throughout the bulk of the atmosphere. At 10 mbar ($\sim$100 km), our upper limits do not exceed a few 10$^{-10}$, consistent with \cites{Vuitton2019} and \cites{Krasnopolsky2014} photochemical models that predict a mixing ratio not exceeding 10$^{-10}$ or a few 10$^{-9}$ at this altitude, respectively.

\subsection{Evolution of polar regions}
%In the middle atmosphere of Titan, polar regions are strongly impacted by the global circulation in winter \citep{Lebonnois2012}. At the winter pole, the upper branch of the circulation air cell subsides leading to an adiabatic heating at high altitudes (300-400 km) and an enrichment in photochemical compounds that are formed at upper altitudes (> 600 km). We discuss below our results on the seasonal changes on temperature and abundance profiles at polar latitudes.

\subsubsection{Northern polar region}
In February 2005, the temperature profile in the northern polar region is warmer between 0.3 mbar ($\sim$230 km) and 3$\times$10$^{-3}$ mbar ($\sim$480 km) than the temperature profile around the equator in December 2004, which is consistent with previous analyses of CIRS observations \citep{Vinatier2015, Achterberg2011} and GCM predictions (\cite{Newman2011, Lebonnois2012, Lora2015};\cite{VatantdOllone2018}, in prep.). Below the 3$\times$10$^{-3}$ mbar pressure level ($\sim$480 km), we also observed an enrichment in all molecules, except CO$_2$, compared to their abundances around the equator, in agreement with the results of \cite{Vinatier2015}, \cite{Teanby2019} and \cite{Sylvestre2018}. This enrichment vastly differs from one compound to another. In the stratosphere, below the 0.1-mbar region, it is most marked for HC$_3$N, C$_6$H$_6$, C$_4$H$_2$, C$_3$H$_4$ and HCN, intermediate for C$_2$H$_2$, and weak for C$_2$H$_6$ and C$_3$H$_8$. From the GCMs predictions, the descending air branch of the pole-to-pole circulation cell brings air enriched in photochemical species from the high atmosphere and heats adiabatically the mesosphere, resulting in a temperature local maximum around 0.01 mbar ($\sim$410 km), instead of 0.1 mbar ($\sim$275 km) for the stratopause in the equatorial region. Below 0.3 mbar ($\sim$230 km), the temperature profile in the northern polar region is colder than around the equator (by $\sim$20 K at 1 mbar, $\sim$175 km), in agreement with the results of \cite{Vinatier2015}, \cite{Achterberg2011} and \cite{Teanby2019}. The smaller stratospheric polar temperatures likely result from the absence of solar heating in the polar night below 300 km. 

%However, the transport of enriched air by the descending branch of the pole-pole circulation cell reached the lower stratosphere, since abundance profiles in the northern polar region are slightly more abundant than those around the equator, consistent with \cite{Vinatier2015} and \cite{Teanby2017} and \cite{Sylvestre2018}.

Between February 2005 and August 2007, we observed a decrease of the temperature of the polar stratopause by $\sim$7 K at 0.01 mbar ($\sim$410 km), consistent with previous CIRS observations \citep{Vinatier2015, Achterberg2011}. This cooling could be due to the weakening of the descending branch of the pole-to-pole circulation cell at the end of winter. This assumption is supported by the decreasing speed of zonal winds around the northern polar region inferred by \cite{Achterberg2011} and predicted by GCMs (\cite{Newman2011, Lebonnois2012, Lora2015}; \cite{Tokano2013}; \cite{VatantdOllone2018}, in prep.). The temperature profiles in the lower stratosphere below 0.3 mbar ($\sim$230 km) did not vary between 2005 and 2007.

From August 2007 to July 2009, the stratopause temperature decreased by $\sim$21 K at 0.01 mbar ($\sim$375 km), which is consistent with previous analyses of CIRS observations \citep{Vinatier2015, Achterberg2011}. This trend is not reproduced by \cites{Newman2011} GCM which predicts warmer temperatures (in excess of 210 K) between 0.1 and 0.01 mbar at the northern spring equinox (11 August 2009) compared with the northern winter solstice (23 October 2002). The \cite{VatantdOllone2018} (in prep.) GCM predicts this decreasing temperature trend, but with an amplitude at 0.01 mbar smaller by 10 K than what we derived. This cooling could be explained by the weakness of the adiabatic heating associated with downwelling motion because of the coexistence of two weaker equator-to-pole circulation cells that is predicted to occur around northern spring equinox (\cite{Newman2011, Lebonnois2012, Lora2015}; \cite{Tokano2013}; \cite{VatantdOllone2018}, in prep.). The lower stratosphere is still colder than at the equator because of the smaller solar flux at low altitudes. In December 2009, we observed an oscillation on the thermal profile that we discuss further in Section \ref{discussion: oscillation}. In April 2010, the retrieved profile shows temperatures between 1 mbar ($\sim$160 km) and 0.01 mbar ($\sim$360 km) lower than in July 2009, but this difference is possibly due to their different latitudes (89 and 67$\degree$). Regarding abundance profiles around the northern spring equinox, we observed a strong enrichment in photochemical compounds above 0.1 mbar ($\sim$250 km) compared to earlier during winter (Fig.\  \ref{fig:north_profqs}). This enrichment in the mesosphere could be due to a combination of the reactivation of the photochemistry in the northern polar region by the increasing solar flux at altitudes higher than 500 km \citep{Lebonnois2001}, and the transport of this enriched air to deeper altitudes by the descending branch of the equator-to-pole circulation cell.

From April 2010 to December 2011, we observed a cooling above $\sim$10$^{-2}$ mbar ($\sim$450 km) and a slight heating below this pressure level consistent with \cite{Vinatier2015} and \cite{Teanby2019}. Moreover, abundance profiles in December 2011 show a depletion in all molecules above 0.1 mbar ($\sim$260 km) compared to April 2010, as previously noted by \cite{Vinatier2015}. This cooling and depletion in the mesosphere could be explained by the establishment of a pole-to-pole circulation cell during the northern spring, in which the ascending branch occurs in the northern polar region, as predicted by GCMs (\cite{Newman2011, Lebonnois2012, Lora2015}; \cite{Tokano2013}; \cite{VatantdOllone2018}, in prep.). In contrast, the lower stratosphere warmed up probably due to the increasing solar flux in the northern polar region. Later in January 2015, the temperature profile had warmed up between 10 mbar ($\sim$100 km) and 0.03 mbar ($\sim$340 km), with an increase of $\sim$9 K at 1 mbar ($\sim$175 km) and $\sim$20 K at 0.1 mbar ($\sim$280 km), consistent with \cites{Coustenis2016} conclusions. This warming is reproduced by \cites{Newman2011} GCM, which predicts a temperature change at 1 mbar between the northern spring equinox and summer solstice consistent with the observations. However, at 0.1 mbar, the observed 20 K decrease is not reproduced by \cites{Newman2011} GCM. The stratospheric heating could be explained by the increasing solar flux reaching the northern polar region. Above 0.03 mbar, the temperature in the mesosphere did not vary between December 2011 and January 2015. In contrast, the abundances of all molecules above 0.1 mbar ($\sim$280 km), except C$_2$H$_6$ and C$_3$H$_8$, decreased over this period, which is consistent with the upwelling provided by the pole-to-pole circulation cell. Actually, Vinatier et al. (in preparation) show that the depleted air is observed down to mid-northern latitudes. 
Surprisingly, the enrichment in photochemical compounds observed in December 2011 between 1 mbar ($\sim$175 km) and 0.1 mbar ($\sim$280 km) was still there in January 2015. This could be explained by the persistence of a circulation cell derived from the northern winter pole-to-pole circulation but confined to the stratosphere and to high northern latitudes as seen in the numerical simulation of  \cite{Lebonnois2012} \cite[see Fig.\ 12 of ][]{Sylvestre2018}.

\subsubsection{Southern polar region}
We discuss here the evolution of the southern polar region starting in 2009 around the northern spring equinox, since there were no limb data acquired at high southern latitudes in the 2005-2008 period, during the northern winter.

In August 2009, the thermal profile in the southern polar region is similar to the one around the equator in March 2009 between 0.3 mbar ($\sim$240 km) and 0.02 mbar ($\sim$370 km), as also derived by \cite{Vinatier2015}, \cite{Achterberg2011}, \cite{Teanby2012} and \cite{Teanby2017}.This similarity is reproduced by \cites{Tokano2013} and \cites{VatantdOllone2018} (in prep.) GCM, while \cites{Newman2011} GCM predicts a $\sim$ 20 K higher temperature in the southern polar region. Below 0.3 mbar ($\sim$240 km), the thermal profile in the southern polar region is colder than the one around the equator (i.e: 5 K at 1 mbar, $\sim$185 km), consistent with the results of \cite{Vinatier2015}, \cite{Achterberg2011}, and \cite{Teanby2019} and roughly with predictions of \cites{Newman2011} GCM and \cites{VatantdOllone2018} (in prep.) GCM. \cites{Tokano2013} GCM does not predict such cooling between the equatorial region and the southern polar region near the northern spring equinox. Above 0.02 mbar ($\sim$370 km), mesospheric temperatures are warmer in the southern polar region than around the equator (by $\sim$10 K at 2$\times$10$^{-3}$ mbar, 480 km). This could result from adiabatic heating in the descending branch of the equator-to-pole circulation cell, as seen in the northern polar region during the northern winter. However, this heating term is weak due to the coexistence of the two equator-to-pole circulation cells predicted by GCMs (\cite{Newman2011, Lebonnois2012, Lora2015}; \cite{Tokano2013}; \cite{VatantdOllone2018}, in prep.). In August 2009, C$_2$H$_2$ and HCN abundance profiles in the southern polar region are enriched above 0.001 mbar ($\sim$520 km) and 0.01 mbar ($\sim$400 km), respectively (Fig.\ \ref{fig:south_profqs}) compared to the equatorial region. These enrichments in the mesosphere are due to the descending branch of the equator-to-pole circulation cell. Later in July 2010, the temperature in the southern polar region above 0.03 mbar ($\sim$455 km) increased by $\sim$10 K, and cooled between 10 mbar ($\sim$95 km) and 0.03 mbar, with a $\sim$8 K decrease at 0.2 mbar ($\sim$255 km). While the mesosphere was probably heated by the adiabatic compression occurring in the descending branch of the equator-to-pole circulation cell, the stratosphere most probably cooled due to the decreasing solar flux. 

In September 2011, the temperature in the southern polar region increased in the mesosphere between 0.08 mbar ($\sim$280 km) and 4$\times$10$^{-3}$ mbar ($\sim$420 km) (by $\sim$7 K at 0.01 mbar, 380 km), while below 0.08 mbar the atmosphere strongly cooled, by $\sim$16 K at 1 mbar (175 km), in agreement with the CIRS data retrievals of \cite{Vinatier2015}, \cite{Coustenis2016} and \cite{Teanby2017}. The increasing temperature in the mesosphere is not reproduced by \cites{VatantdOllone2018} (in prep.) GCM. At this date, all photochemical compound abundance profiles show mesospheric enhancements (see Fig.\  \ref{fig:north_profqs}) compared to abundance profiles in July 2010, consistent with results of \cite{Vinatier2015} and \cite{Teanby2017}. Additionally, \benzene\/ and HC$_3$N appeared for the first time in the southern polar region in September 2011. At this date, this suggests that the pole-to-pole circulation cell is well established in the middle atmosphere, since we observed in the southern polar region enrichment in photochemical compounds and warm temperature in the mesosphere, consistent with \cite{Teanby2017}.

Between September 2011 and March 2015, the atmosphere in the southern polar region strongly cooled between 0.3 mbar (210 km) and 4$\times$10$^{-4}$ mbar (500 km) with a temperature decrease of $\sim$26 K at 0.01 mbar (350 km). This strong cooling of the middle atmosphere is consistent with the results of \cite{Coustenis2019} who inferred a steady decrease of temperature from June 2012 to September 2014 in the 0.01--20 mbar pressure range, and with \cites{Teanby2017} results. GCMs do not predict such a cooling in the mesosphere. Moreover, abundance profiles show a stronger enrichment in molecules above 0.01 mbar ($\sim$350 km) compared to September 2011 (Fig.\  \ref{fig:south_profqs}). These profiles differ from those retrieved by \cite{Teanby2017} that show a local maximum in the C$_2$H$_2$ and C$_3$H$_4$ mole fractions around 440 km that we did not observe. Additionally, our C$_2$H$_2$, HCN, C$_2$H$_6$, HC$_3$N and C$_6$H$_6$ volume mixing ratios at 2$\times$10$^{-5}$ mbar ($\sim$ 645 km) are larger than those retrieved with INMS at 1050 km \citep{Cui2009, Magee2009}. However, \cite{Cui2009} and \cite{Magee2009} have analyzed INMS data during the northern winter (2004-2008) and primarily in the northern hemisphere, so the comparison with CIRS must be taken with caution. According to \cite{Teanby2017}, the enrichment in short lifetime molecules (e.g., C$_3$H$_4$, C$_4$H$_2$, HC$_3$N and C$_6$H$_6$) in the mesosphere strongly increased the efficiency of the long-wave radiative cooling between September 2011 and March 2015, which is consistent with our derived abundance and temperature profiles. This is also consistent with the HCN cloud observed by VIMS in May 2012 at 300 km in the southern polar region that required temperatures close to 125 K to allow HCN to condense \citep{DeKok2014}. A polar cloud was also simultaneously observed by Cassini/ISS \citep{West2016}. In July 2012, CIRS observed an emission feature at 220 \cm\/ attributed to a condensate \citep{Jennings2012}, and in May 2013, \benzene\/ ice was observed below 300 km by \cite{Vinatier2018}.

From March to September 2015, the southern polar atmosphere strongly warmed up between 0.2 mbar (215 km) and 10$^{-4}$ mbar (600 km), by e.g.\ by $\sim$31 K at 0.04 mbar (285 km), as also observed by \cite{Teanby2017}. Over the same period, abundance profiles show a decrease above 10$^{-3}$ mbar (see Fig.\  \ref{fig:south_profqs}), consistent with \cites{Teanby2017} observations. However, the lower stratosphere in the southern polar region is more enriched in trace gases than the equator, in agreement with the results of \cite{Sylvestre2018}. According to \cite{Teanby2017}, the re-appearance of a hot mesosphere is the result of the stabilization of the polar vortex with a slight reduction of the abundance of photochemical compounds, consistent with our observations (Fig.\ \ref{fig:south_profqs}), and the adiabatic heating term outstripping the radiative cooling.

Between September 2015 and January/February 2016, the atmosphere in the southern polar region strongly cooled down below 0.01 mbar ($\sim$ 325 km), e.g.\ by $\sim$30 K at 0.2 mbar (210 km), in agreement with \cites{Teanby2017} results. This stratospheric cooling is likely due to the lack of solar flux in this region and the relative weakness of the adiabatic heating as pressure increases. These thermal profiles show a colder stratopause than those observed in the northern polar region during the winter in 2005. Abundance profiles in January 2016 are similar to those in September 2015 between 0.1 mbar ($\sim$230 km) and 10$^{-4}$ mbar ($\sim$565 km), consistent with \cites{Teanby2017} results, while between January and February 2016 abundance profiles have become relatively constant with altitude.

Between October 2014 and June 2016 at 51$\degree$S, the mesosphere above 0.04 mbar (310 km) significantly warmed up (by $\sim$11 K at 0.01 mbar, 375 km) while the atmosphere below the 0.04-mbar region cooled down by 1-2 K. This mesospheric heating could be explained by an increase of the adiabatic heating in the descending branch of the pole-to-pole circulation cell. \cite{Teanby2017}  suggested that the southern polar vortex was extending to $\sim$60$\degree$S in 2016, which is consistent with our observations.

\subsection{Thermal oscillation in the northern mesosphere near equinox}
\label{discussion: oscillation}
\begin{figure}[h]
\includegraphics[width=\linewidth]{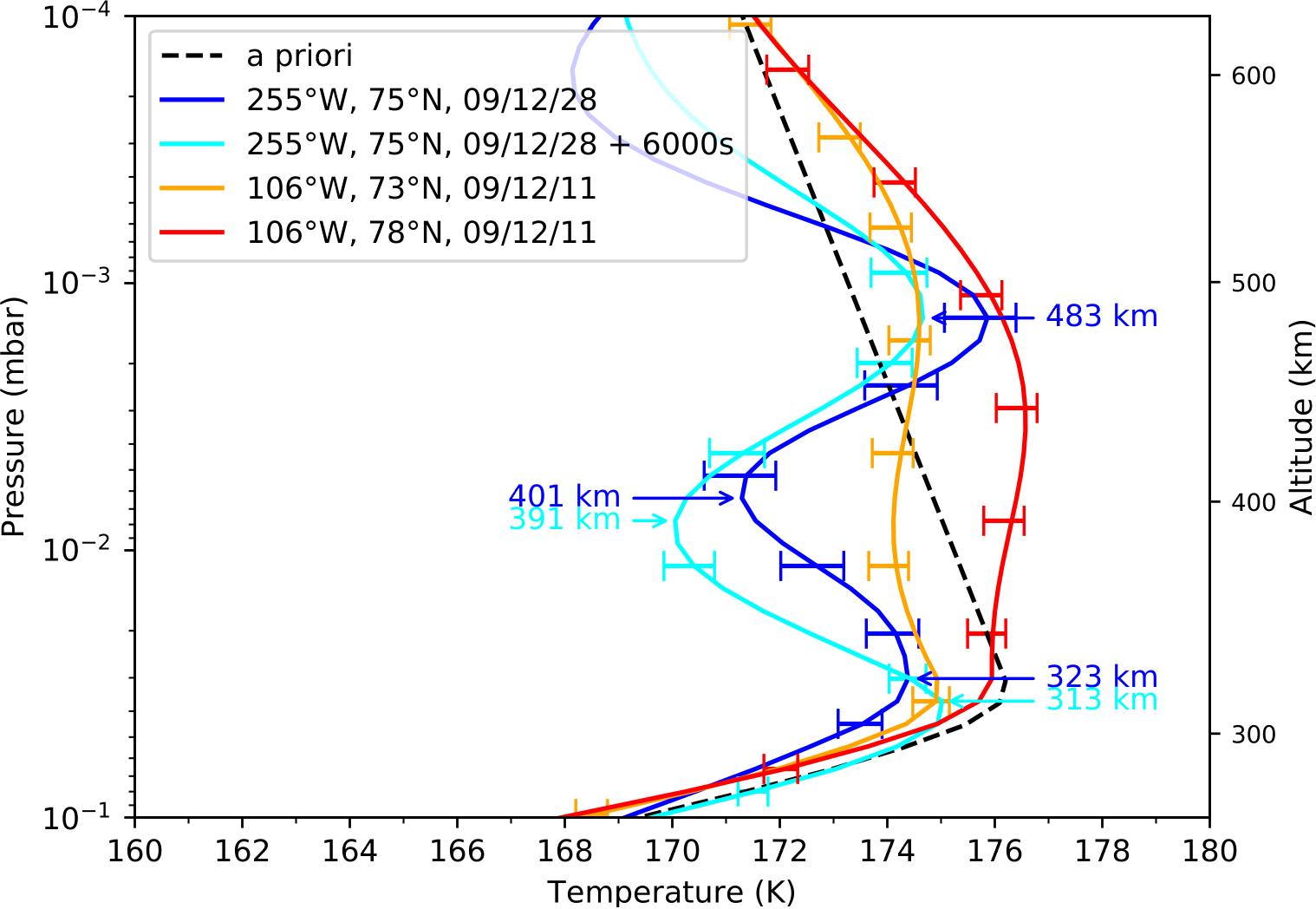}
\caption{Temperature profiles retrieved on 11 and 28 December 2009: blue and cyan thermal profiles retrieved from spectra recorded at 0.5-\cm\/ spectral resolution on 28 December 2009 (local time: 12 a.m.), longitude 255$\degree$W, latitude 75$\degree$N and spaced in time by 6\,000 s respectively; orange and red thermal profiles retrieved from spectra recorded at 15.5-\cm\/ spectral resolution on 11 December 2009 (local time: 9 p.m.), longitude 106$\degree$W, latitude 73$\degree$N and 78$\degree$N respectively. The dashed line is the \textit{a priori} temperature profile used in all retrievals.}
\label{fig:T64_wave}
\end{figure}

%Read Strobel 2005, Teanby 2009, Porco 2005, Hayes 2013
%Planetary waves : \cite{Cui2014} observe density wave at 950 km with a wavelength of 730 km. The gravitational tidal waves induces by Saturn are expected to be observed at equatorial latitudes \citep{Strobel2006}.
The thermal profile at 74$\degree$N in December 2009 shows an oscillation between 5$\times$10$^{-2}$ ($\sim$300 km) and 10$^{-3}$ ($\sim$500 km) mbar with a peak-to-peak amplitude of 3-4 K. We present here a possible explanation of this thermal oscillation. To do so, we re-analyzed the limb spectra recorded on 28 December 2009 (local time: 12 a.m., longitude: 255$\degree$W) at 0.5-\cm\/ spectral resolution. We split the dataset in two subsets separated by around 6\,000 s, to search for a possible vertical displacement between the two retrieved thermal profiles. We also analyzed limb spectra recorded at 15.5-\cm\/ spectral resolution on 11 December 2009 (local time: 9 p.m., longitude: 106$\degree$W) at 73$\degree$N and 78$\degree$N, in order to check if the oscillation was also present at a prior date. From these different datasets of limb spectra, we retrieved thermal profiles with the same methodology and the same \textit{a priori} profile for each retrieval. In Fig.\ \ref{fig:T64_wave}, we show thermal profiles retrieved from limb spectra at 0.5-\cm\/ spectral resolution (blue and cyan) and 15.5-\cm\/ spectral resolution at 73$\degree$N and 78$\degree$N (orange and red respectively). We conclude that the oscillation persisted at least during 6\,000 s on 28 December 2009, while one Titan day before, the atmosphere did not present such an oscillation at a location close in latitude.

% Waves
Let's assume that this oscillation results from an atmospheric wave. From Fig.\ \ref{fig:T64_wave}, we estimate that the wave moved in phase by $\sim$-10 $\pm$~20 km in 6\,000 s (we are confident about the error bar, which corresponds to half of the altitude gap between two limb spectra), giving a phase speed $c_z$ in the range -5 and +1.7 m s$^{-1}$. We can also measure a vertical wavelength $L_z  \sim$170 km between 5$\times$10$^{-2}$ mbar ($\sim$300 km) and 10$^{-3}$ mbar ($\sim$500 km).

% Acoustic : non vitesse de propagation verticale trop rapide 44 m/s
% Inertial : oui 3.9 m/s
% Inertia-gravity : oui 4.4 m/s
% Gravity : non car soit c'est u_0 = 0 soit c'est Latitude = 0°
% Lambs : non car propagation juste horizontal
% Surface :  non car B != 0
% Centrifugal : non car verifie pas un critere
We determined the type of wave that verifies the above characteristics and fulfills the required criteria among six different types of waves in a cyclostrophic regime according to \cite{Peralta2014a} and \cite{Peralta2014b}. We found that an inertia-gravity wave is relevant to our case. The general dispersion relation is given by:

\begin{equation}
\omega = \pm \sqrt{\frac{N_B^2 k_x^2 + \xi^2\left(m^2 + \frac{1}{4H_0^2} \right)}{k_x^2+m^2+\frac{1}{4H_0^2}}} + k_x u_0
\label{omega}
\end{equation}

where $N_B$ is the Brunt-Väisälä frequency ($\sim$2.4$\times$10$^{-3}$ s$^{-1}$ at 400 km), $k_x$ is the horizontal wavenumber, $m$ is the vertical wavenumber, $H_0$ is the pressure scale height ($\sim$50 km at 400 km), $u_0$ is the zonal wind (65 $\pm$ 15 m s$^{-1}$ at 74$\degree$N and 400 km as retrieved by \cite{Achterberg2011} in 2009), and $\xi$ is a frequency defined as :

\begin{equation}
\xi^2 = 2\psi \left(\psi - \frac{1}{a + z} \frac{\partial u_0}{\partial \phi}\right)
\end{equation}
where $a$ is Titan's radius, $z$ the altitude (400 km), $\phi$ is the latitude and
$\psi$ is the centrifugal frequency defined as:

\begin{equation}
\psi = \frac{u_0}{a + z} \tan (\phi)
\end{equation}
We evaluated the meridional shear of the zonal wind $\partial u_0 / \partial \phi$ from the map of  \cite{Achterberg2011} showing a -40 m s$^{-1}$ variation from 69$\degree$N to 76$\degree$N in 2009, and found that $\xi \sim$1.7$\times$10$^{-4}$ s$^{-1}$.
We find that for any horizontal wavenumber ($k_x = 2\pi n / L_x$, where $L_x$ is the perimeter of the latitude circle at 74$\degree$N and 400 km, and $n$ a positive integer), one of the two solutions of Eq.~\ref{omega} yields a vertical phase velocity $c_z >$ 7 m s$^{-1}$, which is excluded by our observations. For $n \leq$ 10, the other solution yields $c_z$ between 1.3 and -3 m s$^{-1}$, which is consistent with our observations. For $n > 10$, the solutions are not consistent with our constraints on the phase velocity. Note that for these solutions the vertical group velocity as derived from Eq.~\ref{omega} is positive given that $\xi < N_B$, which means that energy is propagating upwards. Their horizontal phase velocity $c_k$ varies between -87 (for $n$ = 1) and +4 m s$^{-1}$ (for $n$ = 10), smaller than the zonal wind $u_0$ = 65 m s$^{-1}$. For $n \leq$ 6, the wave phase moves westwards. The corresponding period (2$\pi / \omega$) is at least 17 hr. 
These inertia-gravity waves verify the criteria of \cite{Peralta2014a}: (1) $\xi^2$ varies by less than a factor of 2 over a reasonable latitude range around 74$\degree$N, e.g.\ 70$\degree$N and 80$\degree$N, (2) the intrinsic horizontal phase velocity $|\overline{c_k}| = |{c_k - u_0}|$ = $\frac{|\overline{w}|}{k_x}$ is higher than $\lambda_z  \times \partial u_0  / \partial z$ ($\sim$50 m s$^{-1}$) for $n \leq$ 10.

% conclusion
This oscillation could thus be explained by an inertia-gravity wave, but the mechanism needed to produce the wave is unknown. As the wave energy is propagating upwards, it could be generated from convection in the troposphere, possibly due to some instability occurring around equinox. 

If the oscillating structure of the 74$\degree$N temperature profiles is not due to a wave, it could have a radiative origin, such as localized haze layers producing enhanced heating or a photochemically-enriched layer around 10$^{-2}$ mbar ($\sim$300 km) that would enhance the radiative cooling. However, the haze extinction profile we simultaneously retrieved did not present significant variations that could be linked to detached haze layers. Also, the molecular abundance profiles we retrieved are relatively uniform around 10$^{-2}$ mbar ($\sim$300 km) where the local minimum in temperature is located. We therefore discard a purely radiative origin for the observed oscillation.

%============================================================================%
% PART: CONCLUSION
%============================================================================%
\section{Conclusion}
The CIRS limb observations acquired at 0.5-\cm\/ spectral resolution have allowed us to monitor seasonal changes of temperature and abundance profiles in the middle atmosphere of Titan from 2004 to 2017, i.e.\ over almost half a Saturnian year. Using the latest calibration version and a new methodology to average limb spectra, we got a better signal-to-noise ratio on limb spectra that allowed us to probe higher altitudes ($\sim$ 650km) than previous CIRS analyses \citep{Vinatier2015, Teanby2017}. Our study is consistent with \cites{Vinatier2015} work between 2006-2012 and \cites{Coustenis2019}, and \cites{Teanby2019} results inferred from analysis of CIRS nadir spectra. All vertical profiles displayed in this paper are available as supplementary material and can be downloaded from the VESPA data portal (Virtual European Solar and Planetary Access, \href{http://vespa.obspm.fr/planetary/data/}{http://vespa.obspm.fr/planetary/data/}). We summarize below some important results:
\begin{itemize}

% pole Nord
\item During the northern winter, in the northern polar region, we observed temperatures in the mesosphere above 0.1 mbar ($\sim$175 km) that are much larger than at equatorial latitudes in the period 2005--2007 (northern winter). This is likely due to compressional heating occurring in the descending branch of a global circulation cell. It results in a stratopause located around 0.01 mbar ($\sim$410 km), one pressure decade above that around the equator (0.1 mbar, 275 km). Downwelling also explains the large enrichment in most photochemical compounds compared to equatorial and mid-latitudes, particularly in the stratosphere below the 0.1-mbar region. The enrichment varies from one species to another, being huge for molecules like HC$_3$N, C$_6$H$_6$, C$_4$H$_2$, C$_3$H$_4$ and HCN, and weak to inexistent for C$_2$H$_6$, C$_3$H$_8$, and CO$_2$.  In 2009-2010, shortly after the equinox, we derived a strong enrichment in all molecules above the 0.1-mbar ($\sim$250 km) region compared to winter conditions, a possible consequence of the increasing solar flux and local photochemical production. Simultaneously, the atmosphere cooled in the whole 1-0.001 mbar region (165-495 km). A year or so later, in December 2011, the atmosphere had cooled above 0.03 mbar ($\sim$320 km) but warmed up below 0.03 mbar, probably due to the increase of solar flux. Till January 2015, temperature continued to increase below 0.03 mbar (340 km), yielding a stratopause located at 0.1 mbar ($\sim$280 km), similar to its location in the equatorial region.

% cellule residuelle en printemps nord au nord
\item During northern spring at high northern latitudes, the enrichment in all photochemical species in the range 1--0.1 mbar (175-280 km) persisted from December 2011 to January 2015, possibly due to a residual circulation cell as predicted by \cites{Lebonnois2012} General Circulation Model \citep[see Fig.\ 12 of] [] {Sylvestre2018}.

% pole sud
\item The southern polar region underwent strong seasonal changes in temperature and abundance profiles from July 2010 to January 2016. From July 2010 to September 2011, the atmosphere below 0.1 mbar ($\sim$280 km) cooled down, likely in response to the decreasing solar flux. From September 2011 to March 2015, the atmosphere in the region 0.2--4$\times$10$^{-4}$ mbar (225--500 km) strongly cooled down, at least partly because of the strong enrichment in photochemical compounds above 0.01 mbar ($\sim$350 km), which increases the efficiency of the radiative cooling \citep{Teanby2017}. Additionally, the mixing ratios we retrieved in March 2015 are larger at 2$\times$10$^{-5}$ mbar ($\sim$645 km) than those measured {\em in situ} with the INMS at 1050 km \citep{Cui2009, Magee2009}. Later, in September 2015, we observed that the temperature had increased between 0.2 mbar (215 km) and 10$^{-4}$ mbar (600 km). According to \cite{Teanby2017}, this is a result of the slight reduction in infrared-active compounds and the strengthening of the pole-to-pole circulation, which allows the adiabatic heating of the descending branch to outstrip the thermal emission cooling. In January 2016, the temperatures below 0.01 mbar ($\sim$325 km) are much lower than in September 2015, probably due to the lack of solar flux. In June 2016, we see that the polar vortex, characterized by an enrichment of the photochemical species, was extending equatorward down to $\sim$50$\degree$S.

% equator
\item The equatorial latitudes present weak seasonal variations of thermal and abundance profiles from the northern winter to the northern spring. We note that equatorial CO$_2$ abundance profiles show a local maximum at 1 mbar (188 km) and a local minimum near 0.1 mbar (290 km) that are not predicted by photochemical models \citep{Krasnopolsky2014, Loison2015, Dobrijevic2016, Vuitton2019}. We have been able to detect HC$_3$N at equatorial latitudes at altitudes as high as $\sim$580 km (3$\times$10$^{-4}$ mbar) with an abundance of a few ppmv, consistent with UVIS stellar occultation retrievals at 550 km \citep{Koskinen2011}.

% Thermal oscillations that could be a barotropic wave
\item An oscillation of the thermal profile between 300 and 500 km has been observed in December 2009 at 74$\degree$N. Its characteristics (vertical wavelength and weak vertical phase velocity) are consistent with an inertia-gravity wave, given the estimated zonal wind and meridional shear at this latitude and time.

% GCM divergence
%\item \textcolor{red}{Titan GCMs have improved to reproduce the dynamics in the middle atmosphere through the last decade. However, cooling events fail to reproduce the observed thermal structure, like around the northern spring equinox in the northern polar region or march 2015 in the southern polar region.  I invite the reader to see the recent paper on GCM intercomparison \cite{Lora2019}.}

\end{itemize}

To conclude, we have shown that the descending branch of the circulation cell strongly impacts the thermal structure and abundances of photochemical products at high latitudes in the mesosphere and stratosphere. Comparing the evolution of both poles is difficult since the coverage by the Cassini mission is not symmetric in season. Moreover, the north pole in winter is 1 AU closer to the Sun than the south pole in winter due to Saturn's eccentricity. This asymmetry in the insolation pattern may lead to an asymmetry in the dynamics at high latitudes. Unfortunately, the southern pole, presently in the polar night, cannot be observed from ground-based facilities such as ALMA \citep{Cordiner2018, Thelen2019} but we can still study the evolution of the north pole during the northern summer, in order to compare with the south pole at a similar season.

\section{Acknowledgments}
C.\ M. thanks the ESEP label (n$\degree$2011-LABX-030) for his support. We acknowledge support from the Centre National d'\'Etudes Spatiales (CNES) and the Programme National de Plan\'etologie (INSU/CNRS). We are grateful to the CIRS team for their unflagging involvement in the processing and calibration of the data. C.\ M. also thanks Véronique Vuitton for fruitful discussions.
%============================================================================%
% PART: REFERENCES
%============================================================================%
%%%%%%%%%%%%%%%%%%%%%%%
%% Elsevier bibliography styles
%%%%%%%%%%%%%%%%%%%%%%%
%% To change the style, put a % in front of the second line of the current style and
%% remove the % from the second line of the style you would like to use.
%%%%%%%%%%%%%%%%%%%%%%%
%% Harvard
\clearpage
\section*{Bibliography}
\bibliographystyle{model2-names}\biboptions{authoryear}
%% `Elsevier LaTeX' style
%\bibliographystyle{elsarticle-num}
\bibliography{library}
%\bibliography{library}

\clearpage
\section*{Appendices}
%============================================================================%
% APPENDIX A
%============================================================================%
\subsection*{Appendix A: list of observations}
\begin{landscape}
\begin{longtable}{|l | c | c | c | c | c | c | c |}
\hline
Flyby	&	Date	 &	Solar               & Solar zenith & Local  &	Latitude	&	Vertical	&	Altitude shift applied for each \\
		&			 & longitude ($\degree$)& angle ($\degree$)& time (hh:mm) &	FP3 ($\degree$) &	resolution (km)	& block 1/2/3/4 (km)\\
		&		&	&			&			&		&		 & (see Appendix B)\\\hline
\endhead
\multicolumn{8}{|l|}{\textbf{90$\degree$N--60$\degree$N}}\\\hline
T3     & 14 Feb 2005 & 301 & 108 & 15:40 &  79 & 40 & -10/-10/-10/-10/-10/-10/-10/-10 \\
T4     & 31 Mar 2005 & 303 & 107 & 08:10 &  82 & 10 & 30/4 \\
T19    & 09 Oct 2006 & 324 & 117 & 19:50 &  62 & 40 & 67/50/53/45 \\
T35    & 30 Aug 2007 & 335 &  92 & 16:10 &  69 & 40 & -9/-9/-9/-9 \\
T59    & 24 Jul 2009 & 359 &  68 & 13:12 &  66 & 40 & 23/23 \\
T64    & 28 Dec 2009 &   5 &  73 & 12:00 &  74 & 40 & -3/-10/-8/-12 \\
T67    & 05 Apr 2010 &   8 &  87 &  1:00 &  88 & 40 & 9/2/4/0 \\
T79    & 13 Dec 2011 &  28 &  77 & 17:35 &  79 & 40 & -9/-11/-9/-9 \\
T108   & 12 Jan 2015 &  64 &  47 & 12:45 &  70 & 40 & 13/5/13/13 \\\hline

\multicolumn{8}{|l|}{\textbf{60$\degree$N--20$\degree$N}}\\\hline
T14  & 20 May 2006 & 318 & 148 & 23:33 &  51 & 40 & -19/-26/-19/-26 \\
T16  & 21 Jul 2006 & 321 & 143 & 22:00 &  47 & 40 & 8/1 \\
T24  & 29 Jan 2007 & 328 &  69 & 15:45 &  30 & 40 & 10/0/0/-5 \\
T43  & 12 May 2008 & 344 &  60 & 14:38 &  39 & 40 & 11/11/1 \\
T54  & 05 May 2009 & 357 &  95 & 18:20 &  22 & 10 & -12/-12/-12/-12 \\
T76  & 08 May 2011 &  21 & 110 &  2:20 &  54 & 10 & 18/18/13/13\\
T84  & 06 Jun 2012 &  34 &  77 &  6:17 &  46 & 40 & 10/5/10/5 \\
T90  & 05 Apr 2013 &  44 & 100 &  4:40 &  26 & 40 & -6/-9/-8/-9 \\
T103 & 20 Jul 2014 &  58 &  67 &  7:00 &  30 & 40 & -3/-13/-13/-13 \\
T123 & 27 Sep 2016 &  83 &  45 &  8:33 &  51 & 10 & -3 \\
T125 & 30 Nov 2016 &  85 &  74 &  5:32 &  51 & 40 & 13/13/16/16 \\\hline

\multicolumn{8}{|l|}{\textbf{20$\degree$N--20$\degree$S}}\\\hline
Tb   & 13 Dec 2004 & 299 &  91 &  5:33 & -15 & 40 &  28/33/33/33\\
T18  & 07 Sep 2006 & 322 &  32 &  9:48 & -12 & 20 &  0/0/0/0\\
T21  & 12 Dec 2006 & 326 & 150 & 21:59 &  17 & 40 &  -3/-8/2/-6 \\
T23  & 12 Jan 2007 & 327 &  21 & 11:21 &   4 & 40 &  7/2/7/2\\
T27  & 25 Mar 2007 & 330 &  55 & 15:48 & -21 & 40 & 20/16/23/18\\
T47  & 19 Nov 2008 & 351 &  20 & 10:43 &  -9 & 40 &  8/2/2/-2\\
T49  & 21 Dec 2008 & 352 & 142 & 21:31 &  16 & 40 &  25/25/15\\
T51  & 21 Mar 2009 & 355 & 156 & 22:26 &   0 & 40 &  20/15/25/15\\
T57  & 22 Jun 2009 & 358 & 110 & 19:20 &  -8 & 20 &  7/7/7/12\\
T83  & 21 May 2012 &  34 &  89 & 17:55 &   1 & 40 &  36/33/35/30\\
T92  & 10 Jul 2013 &  47 & 172 & 00:34 & -19 & 40 &  12/8/12/8\\
T95  & 14 Oct 2013 &  50 &  23 & 12:19 &  -2 & 40 &  6/3/6/-2\\
T96  & 01 Dec 2013 &  51 & 140 &  1:30 &  11 & 40 &  31/28/31/31\\
T102 & 18 Jun 2014 &  57 &  99 &  5:04 &  11 & 40 &  -13/-15/-22/-22\\
T121 & 25 Jul 2016 &  81 &  89 &  6:04 &   1 & 40 &  -19/-22/-14/-15\\
S98  & 17 Feb 2017 &  87 & 161 & 23:00 & -10 &115 &  11/16/21/31\\\hline

\multicolumn{8}{|l|}{\textbf{20$\degree$S--60$\degree$S}} \\\hline
T6   & 22 Aug 2005 & 308 &  89 & 20:07 & -57 & 10 & -1\\
T15  & 02 Jul 2006 & 320 &  45 & 13:50 & -54 & 40 & 5/2/5/2\\
T28  & 11 Apr 2007 & 330 &  74 &  6:45 & -31 & 30 & 4\\
T39  & 21 Dec 2007 & 339 &  99 &  4:30 & -46 & 40 & 13/7/12/4\\
T42  & 25 Mar 2008 & 343 &  92 &  4:55 & -57 & 40 & 1/-9\\
T58  & 24 Jul 2009 & 359 & 108 & 20:25 & -58 & 20 & 0/6/6/6\\
T82  & 18 Feb 2012 &  31 & 141 &  1:54 & -46 & 40 & 1/-9/1/1\\
T101 & 17 May 2014 &  56 & 124 & 19:47 & -34 & 40 & 23/18/23/21\\
T106 & 23 Oct 2014 &  61 & 124 & 19:40 & -44 & 40 & -27/-33/-24\\
T120 & 07 Jun 2016 &  79 & 137 &  3:07 & -49 & 40 & 2/-5/-3/-4\\
T126 & 23 May 2017 &  90 & 128 & 20:00 & -49 & 40 & 10/-10\\\hline

\multicolumn{8}{|l|}{\textbf{60$\degree$S--90$\degree$S}}\\\hline
T61  & 25 Aug 2009 &   1 & 102 &  3:42 & -70 & 40 & 33/23\\
T62  & 12 Oct 2009 &   2 &  88 & 17:28 & -72 & 20 & 5/5/10/5\\
T65  & 12 Jan 2010 &   5 & 108 & 23:34 & -76 & 40 & -5/-7/-5/-7\\
T71  & 06 Jul 2010 &  11 & 103 &  2:15 & -80 & 40 & -19\\
T78  & 11 Sep 2011 &  25 &  98 & 14:40 & -84 & 40 & -1/-8/-4/-8\\
T110 & 16 Mar 2015 &  66 & 122 &  2:37 & -81 & 40 & 108/108/108/82\\
T113 & 29 Sep 2015 &  72 & 113 & 15:40 & -84 & 40 & 71/74/89\\
T115 & 31 Jan 2016 &  76 & 118 &  4:00 & -84 & 40 & -48/-18/-48/12\\
T117 & 16 Feb 2016 &  76 & 126 &  0:00 & -80 & 40 & -29/-39\\\hline
\caption{List of limb observations used in this study.}
\label{table:list_obs}
\end{longtable}
\end{landscape}

\clearpage
%============================================================================%
% APPENDIX B
%============================================================================%
\subsection*{Appendix B: Limb spectra averaging}

\begin{figure}[!h]
\includegraphics[width=\linewidth]{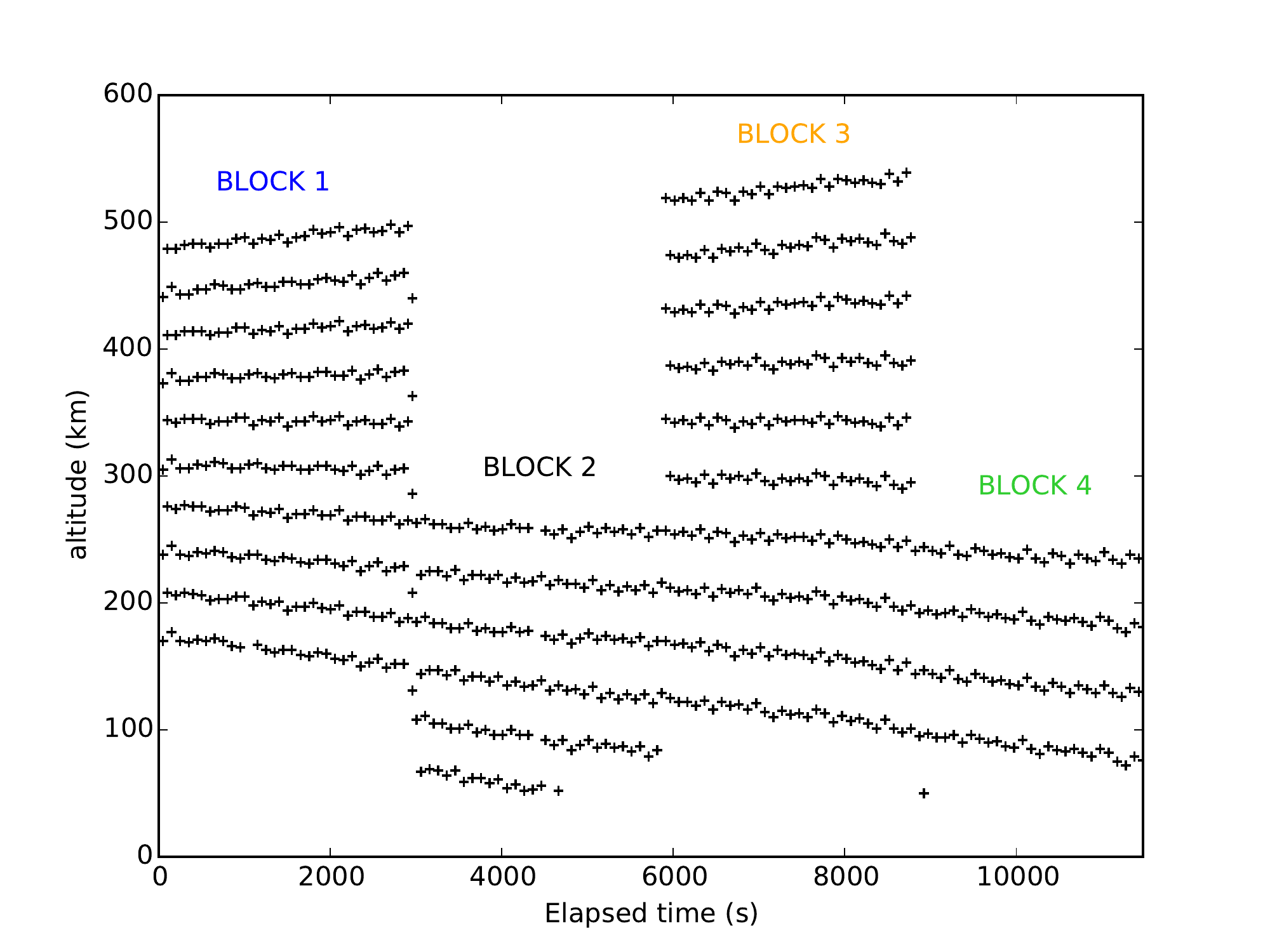}
\caption{Diagram of measurements performed by CIRS during the T64 flyby (Dec 2009, 74 $\degree$N). Each cross corresponds to a spectrum acquired at a given altitude (y-axis) and time (x-axis).}
\label{fig:diagram_block}
\end{figure}

During a given flyby, the CIRS instrument acquired a set of limb spectra in different altitude ranges during sequences that we call ``blocks'' (Fig.\ \ref{fig:diagram_block}). In a block, each of the ten detectors recorded around 30 spectra at roughly the same altitude level, as the spacecraft was approaching or getting away from Titan. For each block, we first averaged spectra detector-by-detector.

\begin{figure}[h]
\includegraphics[width=\linewidth]{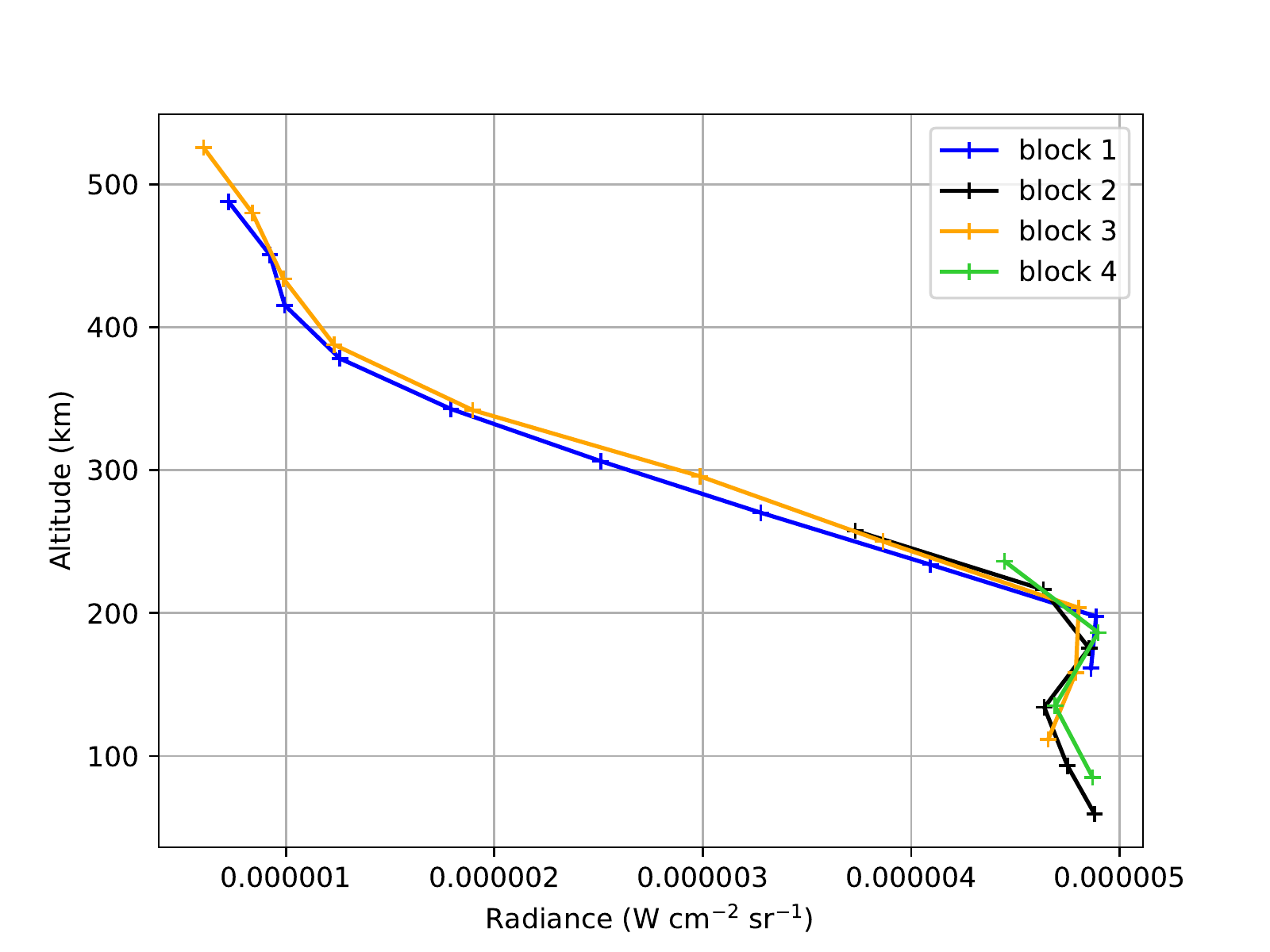}
\caption{Profiles of spectral radiance integrated over the spectral range 1250-1350 cm$^{-1}$ for each block of data shown in Fig.\ \ref{fig:diagram_block}. Each point corresponds to an averaged spectrum for one specific detector. Colors represent the different blocks used.}
\label{fig:rad_init}
\end{figure}

\begin{figure}[h]
\includegraphics[width=\linewidth]{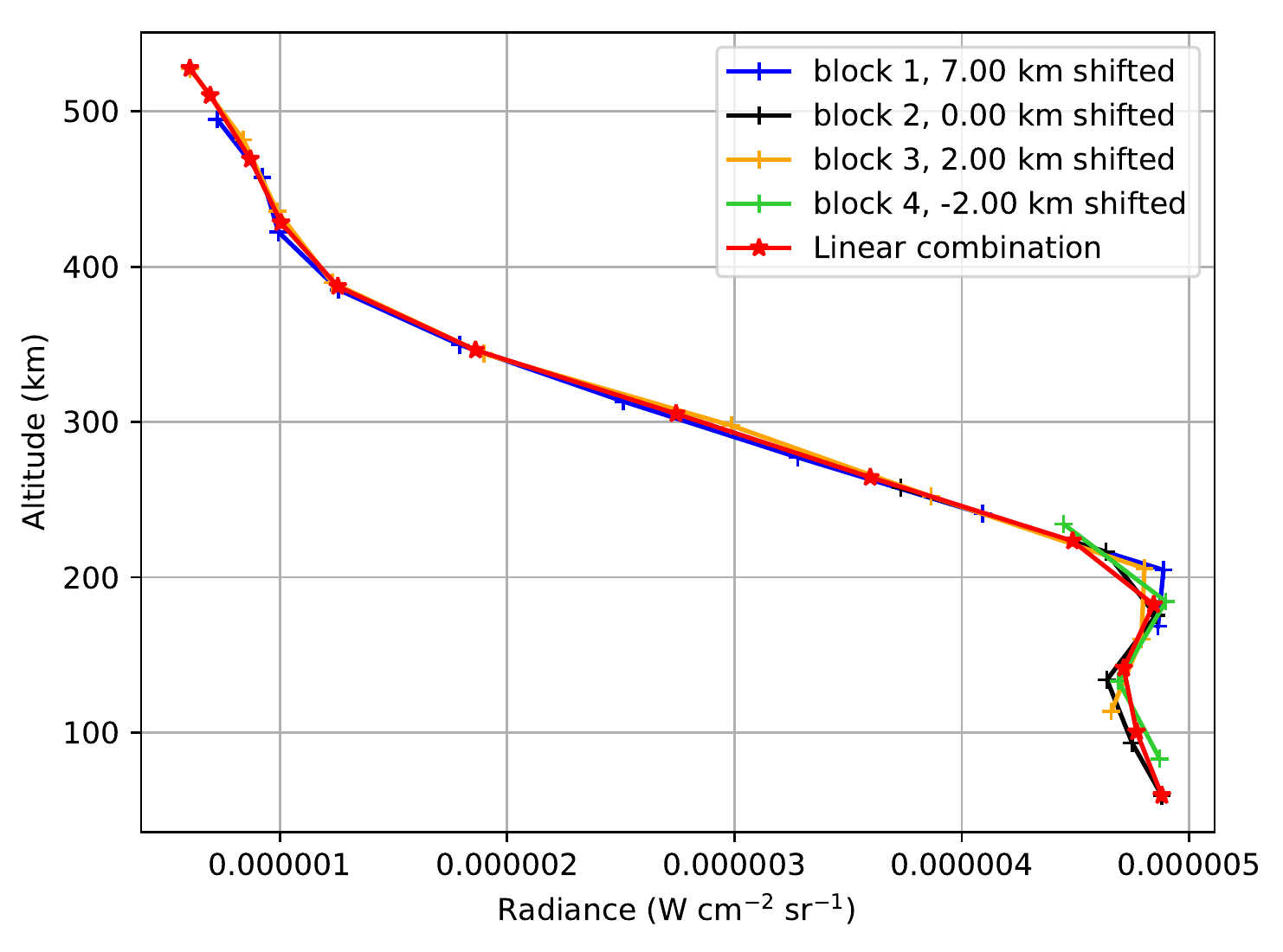}
\caption{Profiles of spectral radiance integrated over the spectral range 1250-1350 cm$^{-1}$ after shifting Blocks 1, 3, 4 relatively to Block 2. Each point corresponds to an averaged spectrum for one specific detector. Colors represent the different blocks used. The red profile correspond to the linear combination of the radiances from all four blocks.}
\label{fig:rad_final}
\end{figure}

We considered the vertical profile of the spectral radiance in each block, using the ten (or less) available averaged spectra (Fig.\ \ref{fig:rad_init}). We integrated the radiance over the spectral range 1250-1350 cm$^{-1}$ of the $\nu_4$ band of methane, and altitudes were extracted from the CIRS database. Without instrument pointing error, we should observe a continuity of the radiance profile from one block to another. In fact, we noticed that, for some flybys, we needed to apply a relative altitude shift from one block to another. We used one of the two deepest blocks as a reference block (e.g.\ Block 2 in Fig.\ \ref{fig:diagram_block}). The reference block is chosen based upon two criteria: (1) a medium vertical resolution and (2) one of the two deepest blocks since we determined the vertical shift from the fit of the P- and Q-branches of the $\nu_4$ methane band in the deepest limb spectra (Section  \ref{methodology: thermal}). Vertical shift adjustments relative to the pointing information in the CIRS database may originate from the pointing precision achieved through the momentum wheels of Cassini, which is about 4$\times$10$^{-5}$ rad \citep{Flasar2004}. For the limb observations used here, the spacecraft was usually at typically 150\,000 km from Titan, which gives an error on altitude of about 6 km. In the case shown in Fig.\ \ref{fig:rad_final}, we needed to apply relative altitude shifts of about +7 km on Block 1, +2 km on Block 3 and -2 km on Block 4. These relative altitude shifts applied to the blocks are consistent with the spacecraft pointing precision. This is also the case for most of the blocks of limb spectra listed in Table \ref{table:list_obs} (last column). The total altitude shift of each block is the sum of this relative altitude shift and the shift derived from the fit of the P- and Q-branches of the $\nu_4$ methane band described in section \ref{methodology: thermal}. Based on the altitude of limb spectra in the reference block, we built an altitude grid with a step equal to the vertical resolution of the observation. Taking into account the relative altitude shift of each block, we linearly interpolated the spectra of each block over this grid and, at each point of the grid, calculated the mean of the interpolated values.

\clearpage
%============================================================================%
% APPENDIX C
%============================================================================%
\subsection*{Appendix C: negative continuum correction}
\begin{figure}[h]
\includegraphics[width=\linewidth]{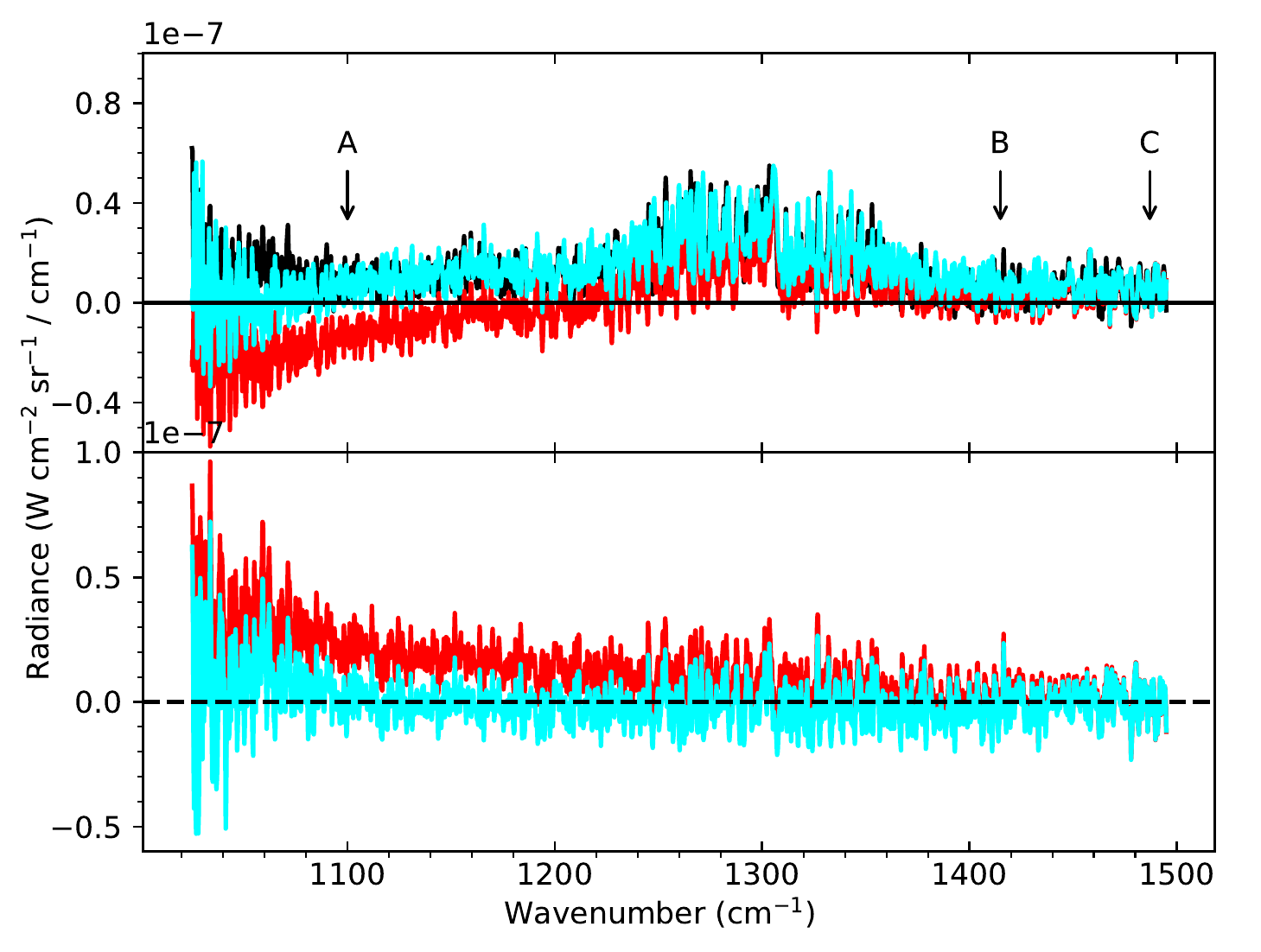}
\caption{Top panel: a spectrum used as a reference (black) and a corrupted spectrum (red) extracted from Detector 30 in Block 1 of Flyby T71. After the correction procedure, we obtain the spectrum shown in blue. Bottom panel: residuals between {\em i)} the reference spectrum and the corrupted spectrum (red), and {\em ii)}  the reference spectrum and the corrected spectrum (blue).}
\label{fig: negative_continuum}
\end{figure}

Some calibrated FP4 limb spectra show a negative continuum from 1075 to 1200 cm$^{-1}$ at high altitudes ($\gtrsim$ 450 km) (see Fig.\ \ref{fig: negative_continuum} in red) where the signal is low (SNR $<$ 1). When selecting spectra from the database, we removed from the average the spectra showing by eye a negative continuum, referred hereafter as ``corrupted'' spectra. However, the averaged spectra we obtained still presented in a few cases a negative continuum, which leads to an underestimation of the retrieved temperature profile. This problem is mostly visible on T21 and T23 FP4 limb spectra. We tried an approach to correct the continuum in such averaged spectra and estimate the error propagation to the retrieved temperature profile.

In the simple case shown here, we extracted from the same block and detector two adjacent spectra: a correct spectrum (used as a reference), and a corrupted one (Fig.\ \ref{fig: negative_continuum}). The difference spectrum is close to 0 at 1500 \cm\ and shows a linear dependence with wavenumber. We therefore decided to add a linear radiance contribution to the corrupted spectrum. To do so, we defined on the corrupted spectrum three points. Point A, at x$_A$ = 1100 cm$^{-1}$, is located in the spectral region that we usually use to retrieve the haze optical depth from the fit of the continuum. Radiance y$_A$ corresponds to the radiance averaged between 1080 and 1120 cm$^{-1}$. Point B is located at x$_B$ = 1415 cm$^{-1}$, between the R branch of the CH$_4$ $\nu_4$ band and the C$_2$H$_6$ $\nu_7$ band, and y$_B$ corresponds to the averaged radiance between 1406 and 1424 cm$^{-1}$. Point C is located at x$_C$ = 1487 cm$^{-1}$, close to the edge of the spectrum.

The linear radiance correction $\delta$y that we added to the corrupted spectrum is:
 
\begin{equation}
\label{corr}
\delta y = \frac{y_A - y_B}{x_B - x_A} * (x - x_C)
\end{equation}

The corrected spectrum is then:

\begin{equation}
I_{corrected} = I_{corrupted} + \delta y
\end{equation}

We added the radiance correction from Eq.\ \ref{corr} to any averaged spectrum in which the continuum appeared to be negative beyond the noise level (only for flyby T21 and T23). This empirical procedure satisfactorily removes the negative continuum but may be not optimal as the distortion of the continuum may differ from the linear trend given by Eq.\ \ref{corr} which was based on a single case.

\clearpage
%============================================================================%
% APPENDIX D
%============================================================================%
\subsection{Appendix D: Influence of the methane abundance on the retrievals}
\label{discussion: methane}

Our hypothesis of a constant methane mixing ratio in the stratosphere at all latitudes may be questioned, considering the work of \cite{Lellouch2014}. Analyzing Cassini/CIRS spectra of both the rotation and $\nu_4$ methane bands, these authors derived latitudinal variations of the stratospheric CH$_4$ mixing ratio during the northern winter, ranging from 1\% at the equator and 50-55$\degree$N and S to 1.5\% at 30-35$\degree$N and S and at the poles. We investigated the influence of using 1\% of methane (instead of 1.48\%) on the temperature and molecular abundance profile retrievals. We selected the T49 flyby observations (December 2008, 16\degree N) and conducted the methodology previously used but with a CH$_4$ mixing ratio of 1\% instead of 1.48\%.

\begin{figure}[h]
\centering
\includegraphics[width=0.8\linewidth]{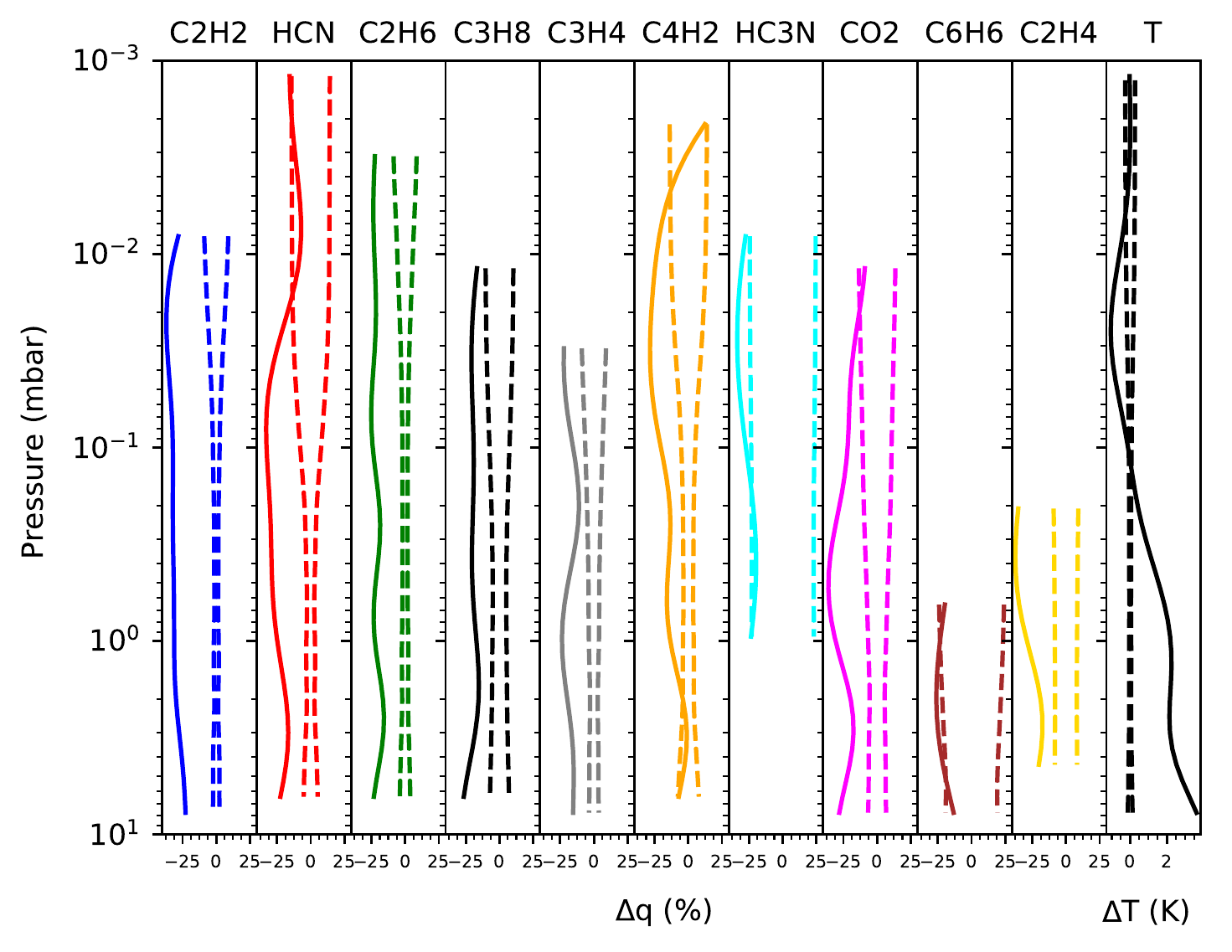}
\caption{Relative variation of the retrieved mixing ratio (\%) and temperature (K) profiles corresponding to a change of the CH$_4$ mixing ratio above the troposphere from 1.48 \% to 1.00\% (solid lines). CIRS spectra recorded in December 2008 at 16$\degree$N were used for this test. Dashed lines correspond to the formal errors attached to the retrieved profiles with 1.48\% of methane.}
\label{fig: methane_1pc}
\end{figure}

%We found a difference in the vertical shift of -7 km with 1\% compared to that found with 1.48\% of methane.
The retrieved temperature profile is at most 4 K warmer with 1\% of methane in the stratosphere below the 0.1-mbar pressure level ($\sim$290 km). This warmer temperature profile impacts the retrieved abundance profiles, which are about 20\% smaller than those retrieved with 1.48\% (Fig.\ \ref{fig: methane_1pc}). We did not take into account this uncertainty on the molecular profiles displayed in Figs.\ \ref{fig:fig_profq_co2}-\ref{fig:fig_profq_hc3n} as \cites{Lellouch2014} analysis is limited to the northern winter and the results may be different during the northern spring. We also chose to be consistent over our whole study in which all retrievals were performed with a uniform CH$_4$ mixing ratio of 1.48\%. This investigation points out the importance of constraining the methane abundance profile as a function of location and season.

\end{document}